\documentclass[12pt,draftclsnofoot,letterpaper,oneside,onecolumn]{IEEEtran}
\def\BibTeX{{\rm B\kern-.05em{\sc i\kern-.025em b}\kern-.08em
    T\kern-.1667em\lower.7ex\hbox{E}\kern-.125emX}}
\usepackage{amsmath}
\usepackage{amsthm}
\usepackage{amsfonts}
\usepackage{amssymb}
\usepackage{graphicx}
\usepackage{cite}
\usepackage{subfigure}
\usepackage{balance}
\usepackage{algorithm}
\usepackage{algorithmic}
\usepackage{enumerate}
\usepackage{color}
\usepackage{array}
\usepackage{indentfirst}
\usepackage{multirow}
\usepackage{booktabs}
\usepackage{color}
\usepackage{slashbox}
\usepackage{diagbox}
\usepackage{threeparttable}
\usepackage[svgnames,table]{xcolor}
\usepackage{bm}
\usepackage{bbm}
\usepackage{subfigure}
\usepackage{epstopdf}

\makeatletter

\newcommand{\Rmnum}[1]{\expandafter\@slowromancap\romannumeral #1@}
\makeatother

\begin{document}
\title{ Deep Residual Learning for Channel Estimation in Intelligent Reflecting Surface-Assisted Multi-User Communications}

\author{\IEEEauthorblockN{Chang Liu, \emph{Member, IEEE}, Xuemeng Liu, \\ Derrick Wing Kwan Ng, \emph{Fellow, IEEE}, and Jinhong Yuan, \emph{Fellow, IEEE} } 

\thanks{C. Liu, D. W. K. Ng, and J. Yuan are with the School of Electrical Engineering and Telecommunications, The University of New South Wales, Sydney, Australia. X. Liu is with The University of Sydney, Sydney, Australia.}

\thanks{This work has been accepted in part to present at the IEEE International Conference on Communications (ICC), 2021 \cite{liu2020deepresidualconference}.}

%

}

%
%

\maketitle

\vspace{-1.6cm}

\begin{abstract}
Channel estimation is one of the main tasks in realizing practical intelligent reflecting surface-assisted multi-user communication (IRS-MUC) systems. However, different from traditional communication systems, an IRS-MUC system generally involves a cascaded channel with a sophisticated statistical distribution. In this case, the optimal minimum mean square error (MMSE) estimator requires the calculation of a multidimensional integration which is intractable to be implemented in practice. To further improve the channel estimation performance, in this paper, we model the channel estimation as a denoising problem and adopt a deep residual learning (DReL) approach to implicitly learn the residual noise for recovering the channel coefficients from the noisy pilot-based observations. To this end, we first develop a versatile DReL-based channel estimation framework where a deep residual network (DRN)-based MMSE estimator is derived in terms of Bayesian philosophy. As a realization of the developed DReL framework, a convolutional neural network (CNN)-based DRN (CDRN) is then proposed for channel estimation in IRS-MUC systems, in which a CNN denoising block equipped with an element-wise subtraction structure is specifically designed to exploit both the spatial features of the noisy channel matrices and the additive nature of the noise simultaneously.
In particular, an explicit expression of the proposed CDRN is derived and analyzed in terms of Bayesian estimation to characterize its properties theoretically.
Finally, simulation results demonstrate that the performance of the proposed method approaches that of the optimal MMSE estimator requiring the availability of the prior probability density function of channel. \vspace{-0.3 cm}
\end{abstract}

\begin{IEEEkeywords}
\vspace{-0.3 cm}
Intelligent reflecting surface (IRS), channel estimation, deep learning, Bayesian estimation.
\end{IEEEkeywords}

\clearpage
\section{Introduction\label{sect: intr}}
Recently, intelligent reflecting surface (IRS), which has the capability of shaping the wireless channels between the users and the base station (BS) to enhance the system performance, has been proposed as a promising technology for the future smart radio environment \cite{gong2020towards, zhao2019survey, liaskos2018new, wong2017key, zhang2020prospective}.
In particular, thanks to the development of advanced materials, a reconfigurable and passive metasurface made of electromagnetic materials has been introduced to design the IRS to make it deployable and sustainable in practice \cite{yang2016programmable, wu2020towards}.
Generally, an IRS is composed of a large number of passive reflecting elements while each element is reconfigurable and can be independently controlled to adapt its phase shift to the actual environment to alter the reflection of the incident signals.
{As such, by jointly adjusting the phase shifts of all the passive elements, a desirable reflection pattern can be obtained which establishes a favourable wireless channel to improve the transmission quality with a low system power consumption \cite{di2019smart, huang2020holographic, alexandropoulos2020reconfigurable}.}
Based on this, an IRS can be adopted in communication systems to enhance the energy efficiency or the spectral efficiency of communication networks through passive beamforming techniques, i.e., designing an efficient configuration of phase shifts to improve the received signal-to-noise ratio (SNR) at the desired receivers \cite{wu2019intelligent}.
Therefore, IRS-assisted communication systems and the related studies such as the network capacity or received SNR maximization \cite{zhang2020capacity, guo2020weighted}, the energy efficiency or spectral efficiency maximization \cite{huang2019reconfigurable, zhou2020spectral}, and the IRS-assisted physical layer security \cite{chu2020intelligent, guan2020intelligent}, etc., have drawn vast attention from both the academia and the industry.

In practice, the promised performance gain brought by an IRS relies on accurate channel state information (CSI) for beamforming. Yet, the aforementioned studies were conducted under the assumption of perfect knowledge of CSI, which is usually not available.
In fact, an indispensable task for realizing IRS-assisted communication systems is to perform accurate channel estimation.
In contrast to the channel estimation in traditional systems, the IRS is passive and cannot perform training sequence transmission/reception or signal processing, i.e., the channel of IRS to user/BS is generally not available separately and only a cascaded channel of user-to-IRS-to-BS can be estimated.
More importantly, the cascaded channel brings two main challenges to IRS-assisted communication systems which are listed as follows:
(i) Limited channel estimation accuracy:
Note that the cascaded user-to-IRS-to-BS channel does not follow the conventional Rayleigh fading model. In this case, the optimal minimum mean square error (MMSE) estimator involves a multidimensional integration which is overly computationally intensive for practical implementation.
Meanwhile, the performance of the available linear MMSE (LMMSE) and least squares (LS) estimators still has a large gap compared with that of the optimal MMSE estimator.
Thus, the channel estimation accuracy is unsatisfactory for practical IRS-assisted communication systems.
(ii) Large channel estimation training overhead:
The IRS generally consists of a large number of elements. Thus, the cascaded channel is with a high dimension and the corresponding channel estimation via conventional methods, e.g., the LS method and the LMMSE method, is computationally costly.

{To overcome the challenge of limited channel estimation performance in IRS-assisted systems (i.e., challenge (i)), a variety of effective algorithms and efficient schemes have been proposed recently.}
For example, in \cite{mishra2019channel}, a binary reflection controlled least squares (LS) channel estimation scheme was developed for single-user systems by switching on only one reflecting element of the IRS and switching off the rest reflecting elements for each time slot. {In this case, the BS only receives interference from the direct link and does not receive any interference from the other reflection elements \cite{yang2020intelligent}. Thus, the BS can estimate the cascaded channel successively, which paves the way for the channel estimation in IRS-assisted systems.}
However, since the binary reflection scheme only active one element each time, only a small received SNR can be obtained at the BS for estimation. Besides, an exceeding long delay may introduce to the system if there is a large number of IRS elements. To further improve the received SNR and shorten the required time for channel estimation, \cite{jensen2020optimal} proposed to switch on all the reflecting elements of the IRS for each time slot. In particular, the authors proposed a discrete Fourier transform (DFT) training sequence-based minimum variance unbiased estimator which can achieve satisfactory estimation accuracy.
{Moreover, the authors in \cite{de2020parafac} developed two parallel factor (PARAFAC)-based methods where the LS estimator was first adopted to obtain coarse channel coefficients and then more accurate estimation results can be obtained through exploiting the least
squares Khatri-Rao factorization and the bilinear estimation techniques.}
On the other hand, some initial attempts have been devoted to the design of efficient schemes to reduce the required training overhead of channel estimation (i.e., challenge (ii)).
For instance, \cite{zheng2020intelligent} and \cite{you2020intelligent} developed a subsurface-based channel estimation scheme where the IRS is divided into several independent subsurfaces and each subsurface is composed of multiple adjacent reflecting elements applying one common phase shift. Hence, by adopting the sharing strategy, the training overhead is dramatically reduced.
Besides, \cite{chen2019channel} introduced a cascaded channel estimation framework and proposed the related algorithm based on the sparse matrix factorization and the matrix completion.
Moreover, \cite{he2019cascaded} studied the channel estimation in IRS-aided multiple-input multiple-output (MIMO) systems.
Specifically, the authors first formulated the channel estimation as a problem of recovering a sparse channel matrix and then proposed a compressed sensing-based scheme to explore the sparsity of the cascaded channel.
Also, in \cite{9120452} and \cite{wang2020channel}, the authors investigated the application of an IRS to a multi-user system and developed a three-phase channel estimation framework to shorten the required time duration for channel estimation by exploiting the redundancy in the reflecting channels of different users.
{Different from the previous works, \cite{alexandropoulos2020hardware} developed a novel IRS architecture equipped with a single active radio frequency (RF) chain for channel estimation at the IRS side, which further shortens the training overhead.}
However, despite the significant research efforts devoted in the aforementioned literature, the problem of limited channel estimation performance in (i) has not been well addressed, thus an effective and practical algorithm which can further improve the estimation accuracy is expected.

{Recently, in contrast to the traditional model-driven approaches, data-driven deep learning (DL) techniques \cite{hinton2006reducing, lecun2015deep, wang2017deep} have proved their effectiveness in IRS-assisted communication systems, such as the DL-based passive beamforming design \cite{taha2019enabling, taha2019deep, huang2019indoor, alexandropoulos2020phase}, the deep reinforcement learning (DRL)-based phase shift optimization \cite{feng2020deep}, and the DRL-based secure wireless communications for IRS-MUC systems \cite{yang2020deep}.} Although these DL-based methods are promising, they still require the availability of perfect CSI for implementation in IRS-assisted systems.
Note that the channel estimation problem is essentially a denoising problem. In particular, the DL technique, especially the deep residual learning (DReL) which adopts a deep residual network (DRN) to improve the network performance and accelerate the network training speed, has been recognized by its powerful capability in denoising in various research areas \cite{he2016deep, zhang2017beyond, he2018deep}.
Motivated by these facts, in this paper, we focus on IRS-assisted multi-user communication (IRS-MUC) systems and adopt a DReL approach to intelligently exploit the channel features to further improve the channel estimation accuracy.
{Note that different from the existing work in \cite{taha2019enabling} where the IRS requires the installation of additional active channel sensors, our work focuses on a general IRS without deploying any active sensors to address the challenging channel estimation problem in IRS-assisted communication systems.
Besides, in contrast to the MMSE channel estimation \cite{neumann2018learning, chen2016adaptive, mirzaei2020mmse}, our proposed method adopts the LS-based channel estimation value as the initial coarse estimated value, models the channel estimation as a denoising problem, and exploits the neural network to design a denoiser.
Thus, our proposed method is model-free and does not require any prior statistical information.
The main contributions of this work are listed as follows{\footnotemark}\footnotetext{
{Note that this paper mainly focuses on improving the estimation accuracy of CSI, but not on reducing the training overhead. As such, the proposed method in this paper adopts the LS-based channel estimation value as the network input and exploits a neural network to further improve the estimation accuracy.
In this case, the training overhead of our proposed method comes from the LS method which can be addressed by adopting the low overhead grouping scheme for IRS elements \cite{zheng2020intelligent, you2020intelligent}.}}:}
\begin{enumerate}[(1)]
\item[(1)] {In contrast to existing channel estimation methods, e.g., \cite{mishra2019channel, jensen2020optimal, de2020parafac, zheng2020intelligent, you2020intelligent, chen2019channel, he2019cascaded, 9120452, wang2020channel, neumann2018learning, chen2016adaptive, mirzaei2020mmse}, we model the channel estimation problem in IRS-MUC systems as a denoising problem and develop a DReL-based channel estimation framework which adopts a DRN to implicitly learn the residual noise for recovering the channel coefficients from the noisy pilot-based observations.}
    Specifically, according to the MMSE criterion, a DRN-based MMSE (DRN-MMSE) estimator is derived in terms of Bayesian philosophy which enables the design of an efficient estimator.

\item[(2)] To realize the developed framework, we adopt a convolutional neural network (CNN) to facilitate the DReL and propose a CNN-based DRN (CDRN) for channel estimation, in which a CNN-based denoising block with an element-wise subtraction structure is specifically designed to exploit both the spatial features of the noisy channel matrices and the additive nature of the noise simultaneously.
    Inheriting from the superiorities of CNN and DReL in feature extraction and denoising, the proposed CDRN method could further improve the estimation accuracy.

\item[(3)] Although it is generally intractable to analyze the performance of a neural network, we formulate the proposed CDRN as a mathematical function and derive an explicit expression of the proposed CDRN estimator to explain the mechanism of the proposed algorithm theoretically, which proves that the proposed CDRN estimator can achieve the same performance as that of the LMMSE estimator.

\item[(4)] Extensive simulations have been conducted to verify the efficiency of the proposed method in terms of the impacts of SNR and channel size on normalized MSE (NMSE), respectively. Additionally, visualizations of the proposed CDRN are also provided to illustrate the denoising process.
    Our results show that the performance of the proposed method approaches that of the optimal MMSE estimator requiring the computation of a prior probability density function (PDF) of cascaded channel.
\end{enumerate}

The reminder of our paper is organized as follows.
Section \Rmnum{2} introduces the system model of the IRS-MUC system and derives the optimal MMSE estimator, the LMMSE estimator, and the LS estimator as benchmarks.
In Section \Rmnum{3}, we model the channel estimation as a denoising problem and develop a versatile DReL-based channel estimation framework.
As a realization of the developed framework, a CDRN architecture and a CDRN-based channel estimation algorithm are designed in Section \Rmnum{4}.
Extensive simulation results are provided in Section \Rmnum{5} to verify the effectiveness of the proposed scheme, and finally Section \Rmnum{6} concludes the work of this paper.

The notations used in our paper are listed as follows.
The superscripts $T$ and $H$ are used to represent the transpose and conjugate transpose, respectively. Terms $\mathbb{R}$ and $\mathbb{C}$ denote the sets of real numbers and complex numbers, respectively. Term $\mathbb{Z}$ denotes the set of integers.
${\mathcal{CN}}( \bm{\mu},\mathbf{\Sigma} )$ is the circularly symmetric complex Gaussian (CSCG) distribution where $\bm{\mu}$ and $\mathbf{\Sigma}$ are the mean vector and the covariance matrix, respectively.
The matrix ${\mathbf{I}}_p$ represents the $p$-by-$p$ identity matrix and the vector ${\mathbf{0}}$ represents a zero vector.
$(\cdot)^{-1}$ denotes the operation of the matrix inverse.
$\mathrm{Re}\{\cdot\}$ and $\mathrm{Im}\{\cdot\}$ are used to represent the operations of extracting the real part and the imaginary part of a complex-valued matrix, respectively.
$\mathrm{tr}[\cdot]$ is the trace of a matrix and $\mathrm{diag}(\cdot)$ represents the construction of a diagonal matrix. $\|\cdot\|_F$ are used to denote the Frobenius norm of a matrix.
In addition, $\exp(\cdot)$ indicates the exponential function, $E(\cdot)$ represents the statistical expectation operation, and $\mathbbm{1}_{\Omega}(\cdot)$ denotes the indicator function of an event $\Omega$.

\begin{figure}[t]
  \centering
  \includegraphics[width=0.6\linewidth]{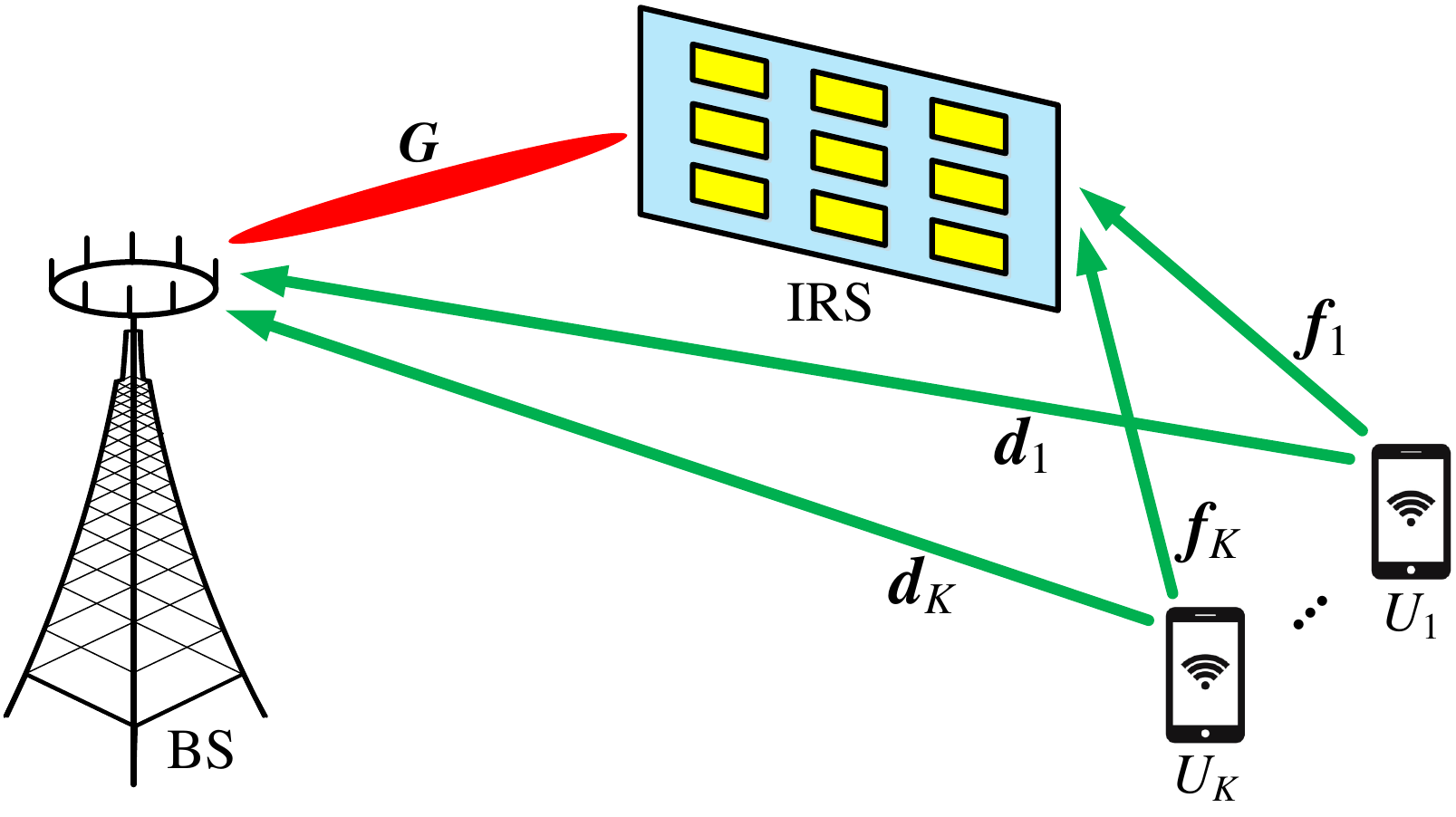} \vspace{-0.8 cm}
  \caption{ The uplink of the considered IRS-MUC system. }\label{Fig:uplink scenario}
  \vspace{-1 cm}
\end{figure}

\section{System Model}
In this paper, we consider an IRS-MUC system adopting the time division duplex (TDD) protocol, which consists of one base station (BS), one IRS, and $K$ users, as shown in Fig. \ref{Fig:uplink scenario}. The BS is equipped with an $M$-element antenna array and $N$ passive reflecting elements are installed at the IRS to assist the BS to serve the $K$ single-antenna users. By joint beamforming design at both the BS and the IRS, the throughput of the IRS-MUC system could be further enhanced. However, the downlink beamforming design critically depends on the availability of accurate CSI. Due to the property of channel reciprocity in TDD systems, the downlink CSI could be acquired from the uplink channel estimation in the IRS-MUC system. The uplink system model is illustrated in Fig. 1, where $U_k$, $k \in \{1,2,\cdots,K\}$, denotes the $k$-th user. The channels of the $U_k$-BS link, the $U_k$-IRS link, and the IRS-BS link are represented by $\mathbf{d}_k \in \mathbb{C}^{M \times 1}$, $\mathbf{f}_k \in \mathbb{C}^{N \times 1}$, and $\mathbf{G} \in \mathbb{C}^{M \times N}$, respectively.
The reflecting link $U_k$-IRS-BS can be regarded as a dyadic backscatter channel \cite{gong2020towards, yang2016programmable}, where each element at IRS combines all the arriving signals and re-scatters them to the BS behaving as a single point source. Denote $\mathbf{R}=\mathrm{diag}(\mathbf{r}) \in \mathbb{C}^{N \times N}$ with $\mathbf{r}=[\beta e^{j\varphi_1},\beta e^{j\varphi_2},\cdots,\beta e^{j\varphi_N}]^T$ as the phase-shift matrix, where $0 \leq \beta \leq 1$ and $0 \leq \varphi_n \leq 2\pi$ are the amplitude and the phase shift at the $n$-th, $n \in \{1,2,\cdots,N\}$, element of IRS, respectively. The channel response from $U_k$ to the BS via the IRS can be expressed as $\mathbf{G}\mathbf{R}\mathbf{f}_k \in \mathbb{C}^{M \times 1}$. In this case, the channel response of the reflecting link $U_k$-IRS-BS can be altered through adjusting the phase-shift matrix $\mathbf{R}$ at the IRS. Since it is obvious that $\mathrm{diag}(\mathbf{r})\mathbf{f}_k=\mathrm{diag}(\mathbf{f}_k)\mathbf{r}$, the channel of reflecting link $U_k$-IRS-BS can be expressed as $\mathbf{G}\mathrm{diag}(\mathbf{r})\mathbf{f}_k = \mathbf{G}\mathrm{diag}(\mathbf{f}_k)\mathbf{r}$. Therefore, the objective of the uplink channel estimation is to estimate
\begin{equation}\label{H_k}
  \mathbf{H}_k = [\mathbf{d}_k,\mathbf{B}_k] \in \mathbb{C}^{M \times (N + 1)}, \forall k,
\end{equation}
where
\begin{equation}\label{}
  \mathbf{B}_k = \mathbf{G}\mathrm{diag}(\mathbf{f}_k) \in \mathbb{C}^{M \times N}, \forall k,
\end{equation}
is a cascaded channel.
Therefore, the downlink channels can be obtained by the conjugate transpose of the uplink channels via exploiting the channel reciprocity in TDD systems.

\begin{figure}[t]
  \centering
  \includegraphics[width=0.6\linewidth]{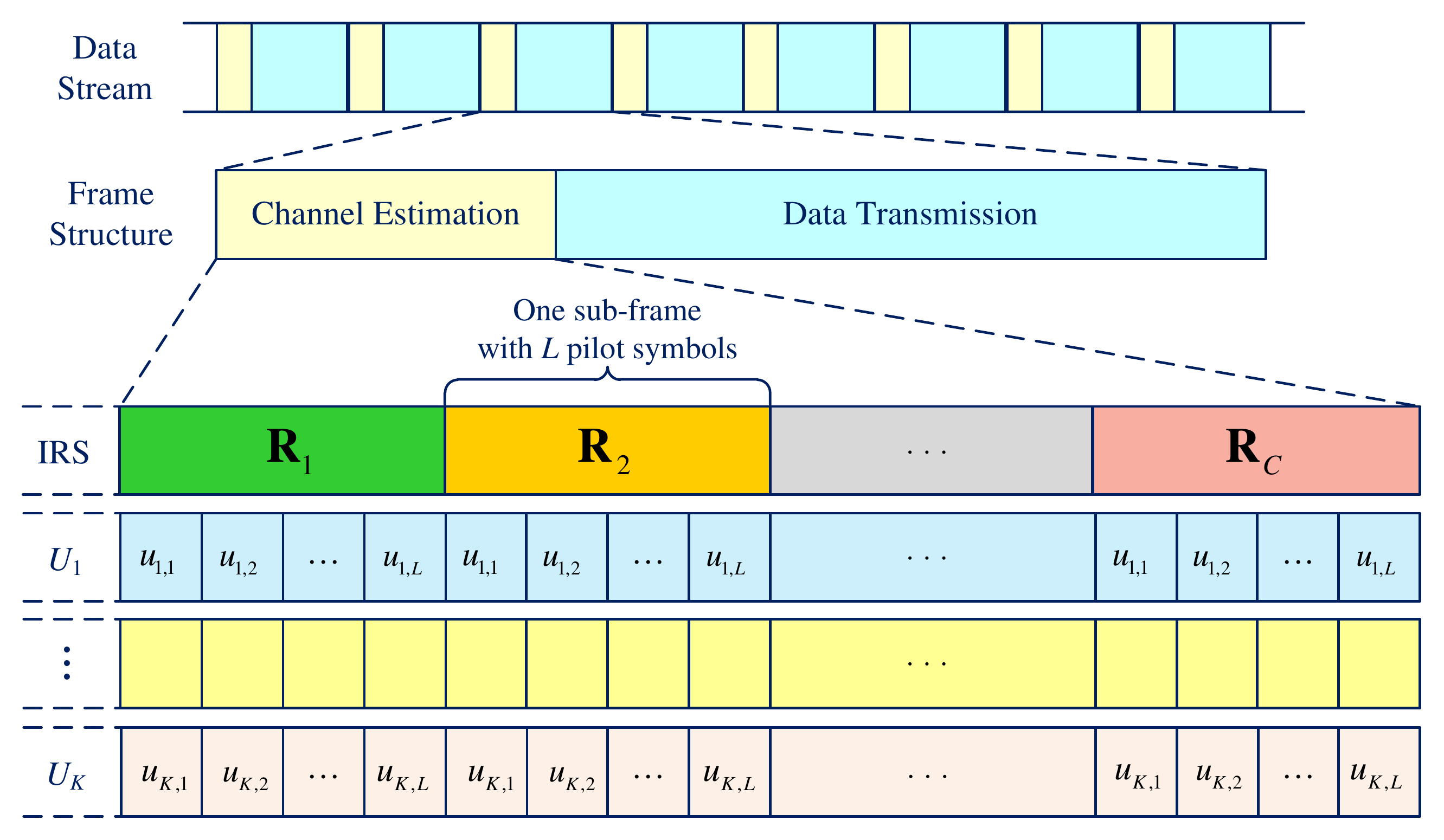} \vspace{-0.6 cm}
  \caption{ Channel estimation protocol for the considered IRS-MUC system. }\label{Fig:channel estimation protocol}
  \vspace{-0.8 cm}
\end{figure}

Based on these, we then develop a channel estimation protocol for the considered IRS-MUC system. As shown in Fig. \ref{Fig:channel estimation protocol}, each frame in the data stream has an identical frame structure which consists of a channel estimation phase and a data transmission phase. In this paper, we mainly focus on the channel estimation phase to estimate $\mathbf{H}_k$ as defined in (\ref{H_k}).
In the channel estimation phase, the IRS generates $C$, $C \geq N + 1$, different reflection patterns through $C$ different phase-shift matrices, denoted by
\begin{equation}\label{Lambda}
  \mathbf{\Lambda} = [\mathbf{R}_1,\mathbf{R}_2,\cdots,\mathbf{R}_C]^{T}.
\end{equation}
Here, $\mathbf{R}_c = \mathrm{diag}(\mathbf{r}_c)$ is the $c$-th, $c \in \{1,2,\cdots,C\}$, phase-shift matrix with $\mathbf{r}_c = [\beta_c e^{j\varphi_{c,1}},\beta_c e^{j\varphi_{c,2}},$ $\cdots,\beta_c e^{j\varphi_{c,N}}]$, where {$\beta_c \in [0,1]$} and $0 \leq \varphi_{c,n} \leq 2\pi$ are the amplitude and the phase shift at the $n$-th element of the IRS under the $c$-th phase-shift matrix, respectively.
In addition, $K$ orthogonal pilot sequences are adopted to distinguish different users, i.e., for $U_k$, $\forall k$, a pilot sequence with a length of $L$, $L \geq K$, is adopted for each IRS phase-shift matrix, denoted by $\mathbf{u}_k = [u_{k,1},u_{k,2},\cdots,u_{k,L}]^{T}$ with $\mathbf{u}_k^{H}\mathbf{u}_k = \mathcal{P}L$ and $\mathbf{u}_a^{H}\mathbf{u}_b = 0$, where $\mathcal{P}$ is the power of each user, $a,b \in \{1,2,\cdots,K\}$, and $a \neq b$.
According to Fig. \ref{Fig:channel estimation protocol}, the channel estimation phase consists of $C$ sub-frames and each sub-frame consists of $L$ pilot symbols. For the IRS, the phase-shift matrix keeps unchanged within one sub-frame and it switches to different phase-shift matrices for different sub-frames. As for the $k$-th user, $U_k$, $\forall k$, it sends its identity pilot sequence $\mathbf{u}_k$ in each sub-frame. Therefore, the $l$-th, $l \in \{1,2,\cdots,L\}$, received pilot signal vector at the BS in the $c$-th sub-frame can be expressed as
\begin{align}
  \mathbf{s}_{c,l} &= \sum\limits_{k = 1}^K (\mathbf{d}_k + \mathbf{G}\mathrm{diag}(\mathbf{r}_c)\mathbf{f}_k){u}_{k,l} + \mathbf{v}_{c,l} \label{x_c(a)} \\
  &= \sum\limits_{k = 1}^K \mathbf{H}_k\mathbf{p}_c{u}_{k,l} + \mathbf{v}_{c,l}.\label{x_c(b)}
\end{align}
Here, $\mathbf{H}_k$ is the channel matrix needed to be estimated as defined in (\ref{H_k}) and $\mathbf{p}_c = [1,\mathbf{r}_c]^T \in \mathbb{C}^{(N + 1) \times 1}$. Equation (\ref{x_c(b)}) is due to the property of $\mathrm{diag}(\mathbf{r}_c)\mathbf{f}_k=\mathrm{diag}(\mathbf{f}_k)\mathbf{r}_c$.
In addition, $\mathbf{v}_{c,l} \in \mathbb{C}^{M \times 1}$ is the $l$-th sampling noise vector at the BS in the $c$-th sub-frame. Generally, $\mathbf{v}_{c,l}$ is assumed to be a CSCG random vector, i.e., $\mathbf{v}_{c,l} \sim \mathcal{CN}(\mathbf{0},\sigma_v^2\mathbf{I}_M)$, where $\sigma_v^2$ denotes the noise variance of each antenna at the BS.
Stacking the $L$ received pilot signal vectors at the BS during the $c$-th sub-frame into a matrix form:
\begin{equation}\label{S_c}
  \mathbf{S}_c = \sum\limits_{k = 1}^K \mathbf{H}_k\mathbf{p}_c\mathbf{u}_k^{H} + \mathbf{V}_c,
\end{equation}
where $\mathbf{S}_c = [\mathbf{s}_{c,1},\mathbf{s}_{c,2},\cdots,\mathbf{s}_{c,L}]$ and $\mathbf{V}_c = [\mathbf{v}_{c,1},\mathbf{v}_{c,2},\cdots,\mathbf{v}_{c,L}]$.
Since the pilot sequences of each two users are orthogonal, the received signal from $U_k$ can be separated by multiplying a sequence $\mathbf{u}_k$ with $\mathbf{S}_c$ in (\ref{S_c}), i.e.,
\begin{equation}\label{x_ck}
   {\mathbf{x}}_{c,k} = \mathbf{H}_k\mathbf{p}_c + {\mathbf{z}}_{c,k}.
\end{equation}
Here, ${\mathbf{x}}_{c,k} = \frac{1}{\mathcal{P}L}\mathbf{S}_c\mathbf{u}_k \in \mathbb{C}^{M \times 1}$ is the received signal vector at the BS from $U_k$ within the $c$-th sub-frame. ${\mathbf{z}}_{c,k} = \frac{1}{\mathcal{P}L}\mathbf{V}_{c}\mathbf{u}_k \in \mathbb{C}^{M \times 1}$ with ${\mathbf{z}}_{c,k} \sim \mathcal{CN}(\mathbf{0},\sigma_z^2\mathbf{I}_M)$, where $\sigma_z^2$ is the variance of each element of ${\mathbf{z}}_{c,k}$.
Based on this and after $C$ subframes, we can then obtain the matrix form of (\ref{x_ck}): $\mathbf{X}_k = [\mathbf{x}_{1,k},\mathbf{x}_{2,k},\cdots,\mathbf{x}_{C,k}]$ which can be expressed as
\begin{equation}\label{X_k}
  \mathbf{X}_k = \mathbf{H}_k\mathbf{P} + \mathbf{Z}_k,
\end{equation}
where $\mathbf{P} = [\mathbf{p}_{1},\mathbf{p}_{2},\cdots,\mathbf{p}_{C}] \in \mathbb{C}^{(N+1) \times C}$ with $\mathbf{p}_c = [1,\mathbf{r}_c]^T$ as defined in (\ref{x_c(b)}) and $\mathbf{Z}_k = [\mathbf{z}_{1,k},\mathbf{z}_{2,k},\cdots,\mathbf{z}_{C,k}] \in \mathbb{C}^{M \times C}$.
According to \cite{jensen2020optimal}, an optimal scheme of $\mathbf{P}$ in terms of improving received signal power at the BS is to design $\mathbf{P}$ as a discrete Fourier transform (DFT), i.e.,
\begin{equation}\label{P}
  \mathbf{P} = \left[
  \begin{matrix}
  1      & 1      & \cdots & 1      \\
 1      & W_C      & \cdots & W_C^{C-1}      \\
 \vdots & \vdots & \ddots & \vdots \\
 1      & W_C^N      & \cdots & W_C^{N(C - 1)}      \\
 \end{matrix}
 \right] \in \mathbb{C}^{(N+1) \times C},
\end{equation}
where $W_C = e^{j2\pi/C}$ and $\mathbf{P}\mathbf{P}^H = C\mathbf{I}_{N + 1}$.
The $C$ different phase-shift matrices $\mathbf{\Lambda} = [\mathbf{R}_1,\mathbf{R}_2,$ $\cdots,\mathbf{R}_C]^{T}$ defined in (\ref{Lambda}) are set based on $\mathbf{P}$ in (\ref{P}).

Therefore, the task of channel estimation in IRS-MUC is to recover $\mathbf{H}_k$, $\forall k$, by exploiting $\mathbf{X}_k$ and $\mathbf{P}$.
Based on the models in (\ref{X_k}) and (\ref{P}), we then first introduce three commonly adopted methods as three benchmarks, namely the LS method, the MMSE method, and the LMMSE method.

(a)~\emph{LS Channel Estimator}:

According to \cite{biguesh2006training}, the estimated $\mathbf{H}_k$ using LS estimator is given by \begin{equation}\label{LSv1}
  \tilde{\mathbf{H}}_k^{\mathrm{LS}} = \mathbf{X}_k\mathbf{P}^\dag,
\end{equation}
where $\mathbf{P}^\dag = \mathbf{P}^H(\mathbf{P}\mathbf{P}^H)^{-1}$ denotes the pseudoinverse of $\mathbf{P}$.

(b)~\emph{MMSE Channel Estimator}:

Different from the LS method where $\mathbf{H}_k$ is assumed to be an unknown but deterministic constant, the Bayesian approach assumes that $\mathbf{H}_k$ is a random variable with a prior PDF $p(\mathbf{H}_k)$.
Therefore, the Bayesian approach can take advantage of prior knowledge to further improve the estimation accuracy.
According to \cite{kay1993fundamentals}, the optimal Bayesian estimator is the MMSE estimator which can be expressed as
\begin{equation}\label{MMSE}
\tilde{ \mathbf{H} }_k^{\mathrm{MMSE}} = E[{\mathbf{H}_k}|{\mathbf{X}_k}] = \int {{\mathbf{H}_k}p({\mathbf{H}_k}|{\mathbf{X}_k})d{\mathbf{H}_k}},
\end{equation}
where $ p({\mathbf{H}_k}|{\mathbf{X}_k}) = \frac{{p({\mathbf{X}_k}|{\mathbf{H}_k})p({\mathbf{H}_k})}}{{\int {p({\mathbf{X}_k}|{\mathbf{H}_k})p({\mathbf{H}_k})d{\mathbf{H}_k}} }}$
is the posterior PDF and $p({\mathbf{X}_k}|{\mathbf{H}_k})$ is the conditional PDF.

(c)~\emph{LMMSE Channel Estimator}:

Note that the optimal MMSE estimator in (\ref{MMSE}) is computationally intensive to be implemented in practice when the channel distribution is complicated.
Thus, the LMMSE estimator can be selected as a more practical approach with a simple explicit expression\footnotemark\footnotetext{{Note that the proposed joint optimization approach in \cite{kang2020intelligent} is an effective scheme for realizing the LMMSE estimation in IRS-assisted systems.
However, the system model in \cite{kang2020intelligent} is a downlink channel estimation model requiring the design of a new training sequence while our work focuses on the uplink channel estimation model for a given training sequence.
Thus, the developed approach in \cite{kang2020intelligent} cannot be directly adopted in our work and the extension of our work to the downlink scenario is left for our future work due to the page limitation.}} \cite{kay1993fundamentals}:
\begin{equation}\label{LMMSE}
  \tilde{\mathbf{H}}_k^{\mathrm{LMMSE}} = \mathbf{X}_k(\mathbf{P}^H\mathbf{R}_{\mathbf{H}_k}\mathbf{P} + M\sigma_z^2\mathbf{I}_C)^{-1}\mathbf{P}^{H}\mathbf{R}_{\mathbf{H}_k},
\end{equation}
where $\mathbf{R}_{\mathbf{H}_k} = E(\mathbf{H}_k^H\mathbf{H}_k)$
denotes the statistical channel correlation matrix \cite{biguesh2006training}.

Note that the LS estimator requires no prior knowledge of channel and is widely used in practice.
If the channel follows the Rayleigh fading model and the statistical channel correlation matrix is available, the optimal MMSE estimator is equivalent to the LMMSE estimator \cite{kay1993fundamentals} and it can be adopted to capture the statistical characteristics of the channel to further improve the estimation performance.
However, for IRS-MUC systems, $\mathbf{H}_k$ includes a cascaded channel which generally does not follow the Rayleigh fading model.
In this case, the optimal MMSE estimator involves the calculation of a multidimensional integration and thus it is intractable to be implemented in practice due to its intensive computational cost.
Meanwhile, the LMMSE estimator still has a performance gap compared with the optimal MMSE estimator \cite{kay1993fundamentals}.

To further improve the estimation performance, in the next section, we still retain the MSE criterion but adopt a data driven DL approach to design a practical channel estimation method, i.e., a DReL-based channel estimation scheme.

\section{Deep Residual Learning-based Channel Estimation Framework}
Note that the additive noise in the system model is a key obstruction for recovering the channel coefficients.
In this section, we first model the channel estimation as a denoising problem and then develop a DReL-based framework to learn the residual noise from the noisy observations for paving the way in recovering the channel coefficients.
In the developed DReL framework, a DRN with a subtraction architecture is designed to exploit both the features of received observations and the additive nature of the noise to further improve the estimation performance.
In the following, we will introduce the derived denoising model and the developed DReL-based channel estimation framework, respectively.

\subsection{Denoising Model for Channel Estimation}
Since a salient feature of the LS estimator is that it does not require any statistical characterization of the data, it is widely adopted in practice due to its convenience of implementation.
According to (\ref{LSv1}), the estimation error of the LS estimator is \cite{biguesh2006training}
\begin{equation}\label{error_LS}
\begin{aligned}
  \varepsilon_{\mathrm{LS}} &= E(\|\mathbf{H}_k - \tilde{\mathbf{H}}_k^{\mathrm{LS}}\|_F^2) \\
  &= M\sigma_z^2\mathrm{tr}[(\mathbf{P}\mathbf{P}^H)^{-1}] \\
  &= \frac{M\sigma_z^2}{(N + 1)C}.
\end{aligned}
\end{equation}
It can be observed that $\varepsilon_{\mathrm{LS}}$ is essentially a function of the noise power and its value decreases with the increase of $C$.

Therefore, if we exploit the LS-based channel estimation value as the initial coarse estimated value, then the channel estimation in the IRS-MUC system can be regarded as a denoising problem:
recovering $\mathbf{H}_k$ from a noisy observation
\begin{equation}\label{LSv2}
  \tilde{\mathbf{X}}_k = {\mathbf{H}}_k + \tilde{\mathbf{Z}}_k,
\end{equation}
where $\tilde{\mathbf{X}}_k = \mathbf{X}_k\mathbf{P}^\dag \in \mathbb{C}^{M \times (N + 1)}$ is the noisy observation based on the LS estimator and $\tilde{\mathbf{Z}}_k = \mathbf{Z}_k\mathbf{P}^\dag \in \mathbb{C}^{M \times (N + 1)}$ denotes the noise component. According to (\ref{LSv2}), $\mathbf{H}_k$ includes a cascaded channel $\mathbf{B}_k$ of which the prior PDF is not a Gaussian PDF, thus, the data model in (\ref{LSv2}) is not a Bayesian general linear model \cite{kay1993fundamentals}.
In this case, it is intractable to derive a Bayesian estimator with an explicit expression as the one adopted in the model driven approach.
In contrast, we will adopt a data-driven approach to develop a DReL-based channel estimation framework for IRS-MUC systems in the following section.

\begin{figure}[t]
  \centering
  \includegraphics[width=\linewidth]{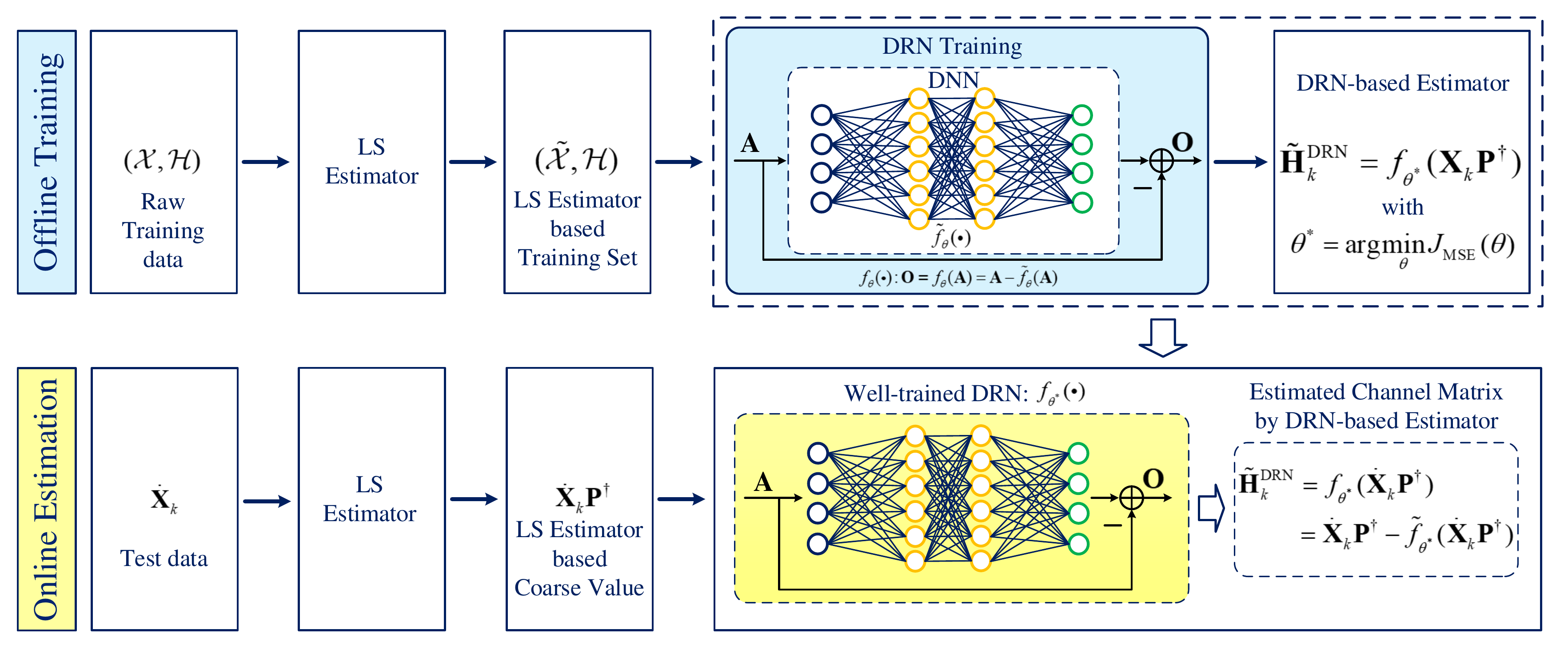} \vspace{-0.6 cm}
  \caption{ The developed DReL based channel estimation framework for IRS-MUC systems.}\label{Fig:DReL_Framework}\vspace{-1cm}
\end{figure}

\subsection{The Developed DReL-based Channel Estimation Framework}
Based on the denoising model, we develop a DReL-based channel estimation framework which consists of an offline training phase and an online estimation phase, as shown in Fig. \ref{Fig:DReL_Framework}.
In the offline training phase, a DRN training process is operated to obtain a well-trained DRN, i.e., the DRN-based channel estimator.
For the online estimation phase, the test data is sent to the well-trained DRN to obtain the estimated channel coefficients directly.
In the following, we will introduce the phases of offline training and online estimation, respectively.

\emph{1)~Offline Training}:

Denote by
\begin{equation}\label{}
  (\mathcal{X},\mathcal{H}) = \{ ( \mathbf{X}_k^{(1)},\mathbf{H}_k^{(1)} ), ( \mathbf{X}_k^{(2)},\mathbf{H}_k^{(2)} ), \cdots, ( \mathbf{X}_k^{(N_t)},\mathbf{H}_k^{(N_t)} )  \}
\end{equation}
the raw training set, where $\mathbf{X}_k^{(i)}$ and $\mathbf{H}_k^{(i)}$ are the input and the ground truth of the $i$-th, $i \in \{1,2,\cdots,N_t\}$, training example, respectively.
To obtain the coarse estimated channel coefficients as defined in the denoising problem of (\ref{LSv2}), we then send $(\mathcal{X},\mathcal{H})$ to the LS estimator in (\ref{LSv1}) and obtain the LS-based training set:
\begin{equation}\label{LS_trainingset}
  (\tilde{\mathcal{X}},\mathcal{H}) = \{ ( \tilde{\mathbf{X}}_k^{(1)},\mathbf{H}_k^{(1)} ), ( \tilde{\mathbf{X}}_k^{(2)},\mathbf{H}_k^{(2)} ), \cdots, ( \tilde{\mathbf{X}}_k^{(N_t)},\mathbf{H}_k^{(N_t)} )  \},
\end{equation}
where $( \tilde{\mathbf{X}}_k^{(i)},\mathbf{H}_k^{(i)} )$ is the $i$-th training example of $(\tilde{\mathcal{X}},\mathcal{H})$.
According to Fig. 3, $(\tilde{\mathcal{X}},\mathcal{H})$ is then sent to DRN for training.
The DRN consists of a DNN and an element-wise subtraction operator between the DNN input and the DNN output. Note that the DNN can be any kind of neural networks, e.g., multi-layer perception \cite{goodfellow2016deep}, CNN \cite{liu2019deep}, etc.
Denote by $\tilde{f}_\theta(\cdot)$ the expression of the DNN where $\theta$ denotes the total parameters of DNN, the output of the DRN can be expressed as
\begin{equation}\label{}
  \mathbf{O} = f_\theta(\mathbf{A}) = \mathbf{A} - \tilde{f}_\theta(\mathbf{A}),
\end{equation}
where $\mathbf{A}$ and $\mathbf{O}$ are the input and output of DRN, and $f_\theta(\cdot)$ denotes the expression of DRN.

In the following, we select the cost function for DRN training.
According to the MMSE criterion, the optimal MMSE estimator is derived by minimizing the Bayesian MSE, i.e.,
\begin{equation}\label{Bmse}
  B_{\mathrm{mse}} = E[\|\tilde{\mathbf{H}}_k - \mathbf{H}_k\|_F^2],
\end{equation}
where $\tilde{\mathbf{H}}_k$ and $\mathbf{H}_k$ denote the estimated channel matrix and the ground truth of the channel matrix.
To achieve the optimal performance, an intuitive selection of the cost function is the Bayesian MSE expression. However, the number of training examples is finite in practice and thus the statistical Bayesian MSE is not available and only the empirical MSE can be adopted which is defined as
\begin{equation}\label{Emse}
  E_{\mathrm{mse}}(N_t) = \frac{1}{2N_t}\sum_{i=1}^{N_t} \left\|\tilde{\mathbf{H}}_k^{(i)} - \mathbf{H}_k^{(i)}\right\|_F^2,
\end{equation}
where $\tilde{\mathbf{H}}_k^{(i)} = f_{\theta}(\tilde{\mathbf{X}}_k^{(i)})$ is the $i$-th estimated channel matrix based on $\tilde{\mathbf{X}}_k^{(i)}$ by the DRN.
In this case, we can only select the empirical MSE as the cost function which can be expressed as
\begin{equation}\label{Jmse}
  J_{\mathrm{MSE}}(\theta) = \frac{1}{2N_t}\sum_{i=1}^{N_t} \left\|f_{\theta}(\tilde{\mathbf{X}}_k^{(i)}) - \mathbf{H}_k^{(i)}\right\|_F^2.
\end{equation}
Based on this, the DRN can adopt the backpropagation (BP) algorithm to obtain a well-trained DRN by progressively updating the network parameters.
Therefore, the well-trained DRN, i.e., the DRN-based MMSE (DRN-MMSE) estimator can be expressed as
\begin{equation}\label{}
  \tilde{\mathbf{H}}_k^{\mathrm{DRN}} = f_{\theta^*}(\tilde{\mathbf{X}}),
\end{equation}
where
$\tilde{\mathbf{X}}$ represents an arbitrary LS-based input and $\theta^*$ denotes the well-trained network parameters by minimizing $J_{\mathrm{MSE}}(\theta)$.

\begin{figure}[t]
  \centering
  \includegraphics[width=\linewidth]{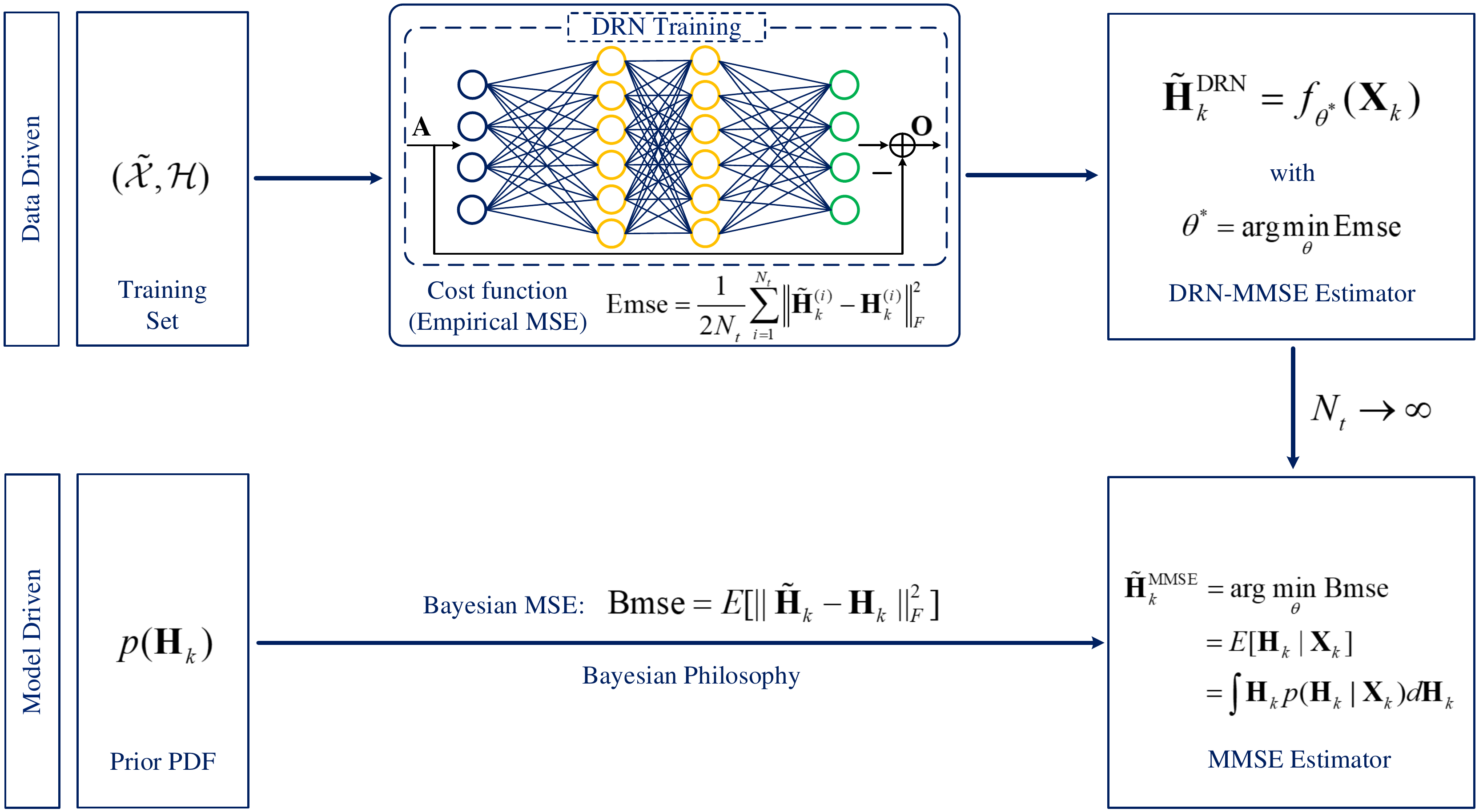} \vspace{-0.6 cm}
  \caption{ A flowchart illustration of the derivation of the DRN-based MMSE estimator. The upper half represents the proposed data driven DReL approach while the lower half is the model driven approach. }\label{Fig:DReL_Analysis}
  \vspace{-1 cm}
\end{figure}

For further understanding, we then analyze the performance of the DRN-based estimator in terms of Bayesian philosophy.
As shown in Fig. \ref{Fig:DReL_Analysis}, there are two approaches to design the MMSE estimator: model driven approach and data driven approach.
Traditionally, given a prior PDF, the MMSE estimator can be derived by minimizing the Bayesian MSE in a model driven manner. However, for the cascaded channel, it is intractable to implement the MMSE estimator due to a lack of closed-form expression and the associated high-cost in computation.
Alternatively, the proposed DRN-MMSE estimator is obtained by minimizing the empirical MSE based on the training set in a data driven approach.
Note that it is obvious that
\begin{equation}\label{}
  \lim_{N_t \to \infty} E_{\mathrm{mse}}(N_t) = \lim_{N_t \to \infty} \frac{1}{2N_t}\sum_{i=1}^{N_t} \left\|\tilde{\mathbf{H}}_k^{(i)} - \mathbf{H}_k^{(i)}\right\|_F^2 = B_{\mathrm{mse}}.
\end{equation}
{Hence, the DRN estimator converges to the MMSE estimator when $N_t \rightarrow \infty$.
As a summary, the DRN estimator is a data driven approach by minimizing the empirical MSE based on a training set while the optimal MMSE is a model driven approach by minimizing Bayesian MSE based on a prior PDF. In particular, the performance of the DRN estimator converges to the optimal MMSE estimator when the training set is sufficiently large.}
Next, we will introduce the online estimation phase of the developed framework.

\emph{2)~Online Estimation}:

Given the test data $\dot{\mathbf{X}}_k$, we first adopt the LS estimator to obtain the initial coarse channel estimation $\dot{\mathbf{X}}_k{\mathbf{P}}^\dag$ and then send it to the well-trained DRN to operate the channel estimation, as shown in Fig. \ref{Fig:DReL_Framework}. Finally, the estimated channel coefficients by the proposed DRN estimator can be expressed as
\begin{equation}\label{}
  \tilde{\mathbf{H}}_k^{\mathrm{DRN}} = f_{\theta^*}({\dot{\mathbf{X}}}{\mathbf{P}}^\dag)={\dot{\mathbf{X}}}{\mathbf{P}}^\dag - \tilde{f}_{\theta^*}({\dot{\mathbf{X}}}{\mathbf{P}}^\dag).
\end{equation}

\textbf{Remark 1}: Note that the DRN in the developed DReL framework can be realized by any kind of DNN architecture, e.g., the dense neural network \cite{goodfellow2016deep} and the CNN \cite{lecun2015deep}.
Therefore, the developed DReL based channel estimation framework is a universal channel estimation framework which can strike a balance between system performance and computational complexity via the selection of DRN.
Specifically, the DRN makes the best use of the DNN with an element-subtraction operator to exploit both the channel features and additive nature of noise to further improve the channel estimation performance for IRS-MUC systems.

\section{CNN-based Deep Residual Network for Channel Estimation}
In this section, we introduce a practical approach for implementing the developed DReL framework via a CNN.
Note that CNN has powerful capability in extracting features from the noisy observation matrices, meanwhile the subtraction structure of DReL contributes to exploiting the additive nature of the noise \cite{he2016deep, zhang2017beyond, he2018deep}.
Therefore, we adopt a CNN to facilitate the application of DReL, resulting a CDRN for channel estimation.
In the following, we will introduce the CDRN architecture and the CDRN-based channel estimation algorithm, respectively.

\subsection{CDRN Architecture}
\begin{figure}[t]
  \centering
  \includegraphics[width=\linewidth,height=0.26\linewidth]{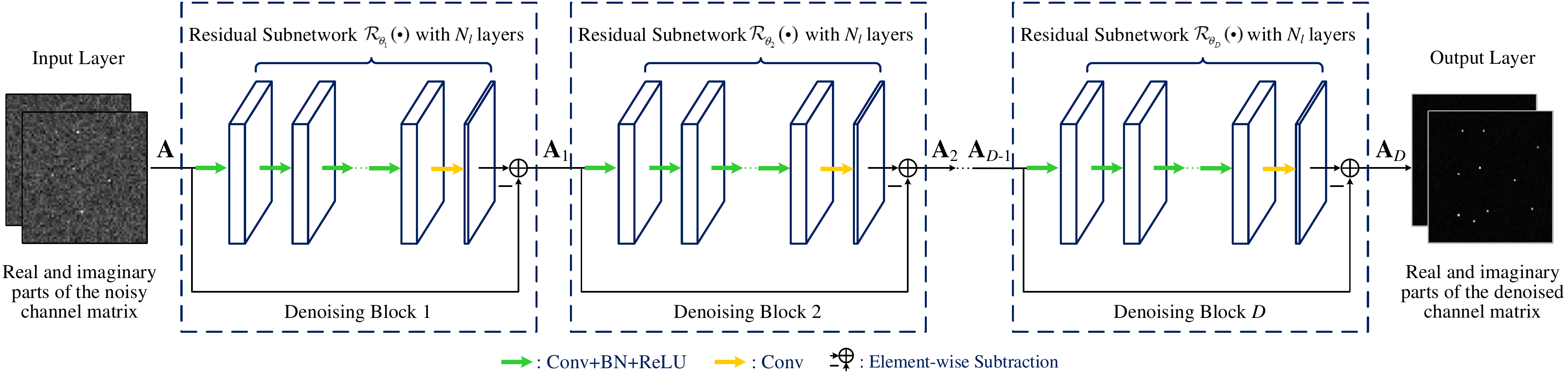} \vspace{-1.6 cm}
  \caption{ The network architecture of the proposed CDRN for channel estimation in the IRS-MUC system. }\label{Fig:CDRN_architecture}
  \vspace{-0.4 cm}
\end{figure}

As shown in Fig. \ref{Fig:CDRN_architecture}, the CDRN consists of one input layer, $D$ denoising blocks, and one output layer.
The denoising block utilizes the CNN with a subtraction structure to learn the residual noise from the noisy channel matrix for denoising.
In addition, considering the channel matrix is a complex-valued matrix, we divide the input into two different parts: the real part and the imaginary part for the convenience of extracting features\footnotemark\footnotetext{{{Since the real part and the imaginary part of the input channel matrix are generally orthogonal \cite{tse2005fundamentals}, we divide the neural network input into two independent parts and adopt the real-valued convolution to exploit features from the real part and the imaginary part of channel matrix, respectively.}}}.
The hyperparameters of CDRN are summarized in Table I and the details of each layer of CDRN are introduced as follows.

\begin{table}[t]
\normalsize
\caption{Hyperparameters of the proposed CDRN}
\vspace{-0.5cm}
\centering
\small
\renewcommand{\arraystretch}{0.95}
\begin{tabular}{c c c}
  \hline
  \vspace{-0.6cm} \\
   \multicolumn{3}{l}{\textbf{Input}:
   {Noisy channel matrix with the size of $M \times (N+1) \times 2$}}  \\
  \hline
  \vspace{-0.6cm} \\
   \multicolumn{3}{l}{\textbf{Denoising Block}:} \\
  \vspace{-0.6cm} \\
  \hspace{0.6cm} \textbf{Layers} & \textbf{Operations} &  \hspace{0.6cm} \textbf{Filter Size}   \\
  \vspace{-0.6cm} \\
  \vspace{-0.6cm} \\
  \hspace{0.6cm} 1 & Conv + BN + ReLU  & \hspace{0.6cm}  $ 64 \times ( 3 \times 3 \times 2 ) $   \\
  \vspace{-0.6cm} \\
  \vspace{-0.6cm} \\
  \hspace{0.6cm} $ 2 \sim N_l-1 $ & Conv + BN + ReLU & \hspace{0.6cm}  $ 64 \times ( 3 \times 3 \times 64 ) $   \\
  \vspace{-0.6cm} \\
  \vspace{-0.6cm} \\
  \hspace{0.6cm} $N_l$ & Conv & \hspace{0.6cm} $ 2 \times ( 3 \times 3 \times 64 ) $  \\
  \vspace{-0.6cm} \\
  \hline
  \vspace{-0.6cm} \\
   \multicolumn{3}{l}{\textbf{Output}:
   {Denoised channel matrix with the size of $M \times (N+1) \times 2$}}  \\
  \vspace{-0.6cm} \\
  \hline
\end{tabular}
\vspace{-1 cm}
\end{table}

\emph{a)~Input Layer}:

To extract features from the complex-valued noisy channel matrix, it is necessary to adopt two neural network channels for the real part and imaginary part of the noisy channel matrix, respectively. Thus, the input layer can be expressed as
\begin{equation}\label{I_input}
  \mathbf{A} = \mathcal{F}([ \mathrm{Re}\{\tilde{\mathbf{X}}_k^{(i)}\}, \mathrm{Im}\{\tilde{\mathbf{X}}_k^{(i)}\} ]),
\end{equation}
where $\mathbf{A} \in \mathbb{R}^{M \times (N+1) \times 2}$ is the input of CDRN, $\mathcal{F}(\cdot):\mathbb{R}^{M \times (2N+2)}\mapsto\mathbb{R}^{M \times (N+1) \times 2}$ represents the mapping function, and $\tilde{\mathbf{X}}_k^{(i)}$ is the LS-based input from an arbitrary training example.

\emph{b)~Denoising Blocks}:

In CDRN, we adopt $D$ denoising blocks to gradually enhance the denoising performance and all denoising blocks have an identical structure.
Specifically, a denoising block consists of a residual subnetwork and an element-wise subtraction operator, as depicted in Fig. \ref{Fig:CDRN_architecture}.
The residual subnetwork of each denoising block has $N_l$ layers.
For the first $N_l-1$ layers, the ``Conv+BN+ReLU'' operations denoted by the green arrows are adopted for each of the first $N_l-1$ layers. The convolution (Conv) operation and the rectified linear unit (ReLU) are adopted together to exploit the spatial features of the channel matrix. To improve the stability of the network and to accelerate the training speed of CDRN, a batch normalization (BN) \cite{ioffe2015batch} is added between Conv and ReLU.
For the last layer, a convolution operation denoted by the yellow arrows is finally adopted to obtain the residual noise matrix for the subsequent element-wise subtraction.
Since the noise in the received signals is of additive nature, an element-wise subtraction is added between the input and the output of the residual subnetwork to obtain the denoised channel matrix.

Denote by $\mathcal{R}_{\theta_d}(\cdot)$ the function expression of the $d$-th, $d \in \{1,2,\cdots,D\}$, residual subnetwork where $\theta_d$ is the network parameters.
The $d$-th denoising block can be expressed as
\begin{equation}\label{I_d}
  \mathbf{A}_d = \mathbf{A}_{d-1} - \mathcal{R}_{\theta_d}(\mathbf{A}_{d-1}), \forall d,
\end{equation}
where $\mathbf{A}_0 = \mathbf{A}$, $\mathbf{A}_{d-1}$ and $\mathbf{A}_d$ are the input and the output of the $d$-th denoising block, respectively.


\emph{c)~Output Layer}:

The output layer is the the denoised channel matrix based on $D$ denoising blocks, i.e., the output of the $D$-th denoising block:
\begin{equation}\label{I_D}
  \mathbf{A}_D = \mathbf{A}_{D-1} - \mathcal{R}_{\theta_D}(\mathbf{A}_{D-1}).
\end{equation}
{According to (\ref{I_d}) and (\ref{I_D}), the output of the CDRN can be finally expressed as
\begin{equation}\label{h_theta}
  g_{\theta}(\mathbf{A}) = \mathbf{A} - \sum\limits_{d = 1}^{D}\mathcal{R}_{\theta_d}\left(\mathbf{A}_{d-1}\right),
\end{equation}
where $g_{\theta}(\cdot)$ denotes the function expression of CDRN with parameters $\theta = [\theta_1,\theta_2,\cdots,\theta_D]$ and $\sum_{d = 1}^{D}\mathcal{R}_{\theta_d}\left(\mathbf{A}_{d-1}\right)$ represents the residual noise component.}
Therefore, the denoised channel matrix is the result of the element-wise subtraction between the noisy channel matrix $\mathbf{A}$ and the residual noise component $\sum_{d = 1}^{D}\mathcal{R}_{\theta_d}\left(\mathbf{A}_{d-1}\right)$.

\textbf{Remark 2}: Note that the neural network is of powerful scalability \cite{goodfellow2016deep}. Although the channel size changes with the settings of the IRS-MUC systems, the proposed CDRN architecture is still scalable and can be easily extended to different shapes accordingly, as will be shown in the simulation results of Section \Rmnum{5}.

\begin{figure}[t]
  \centering
  \includegraphics[width=0.96\linewidth]{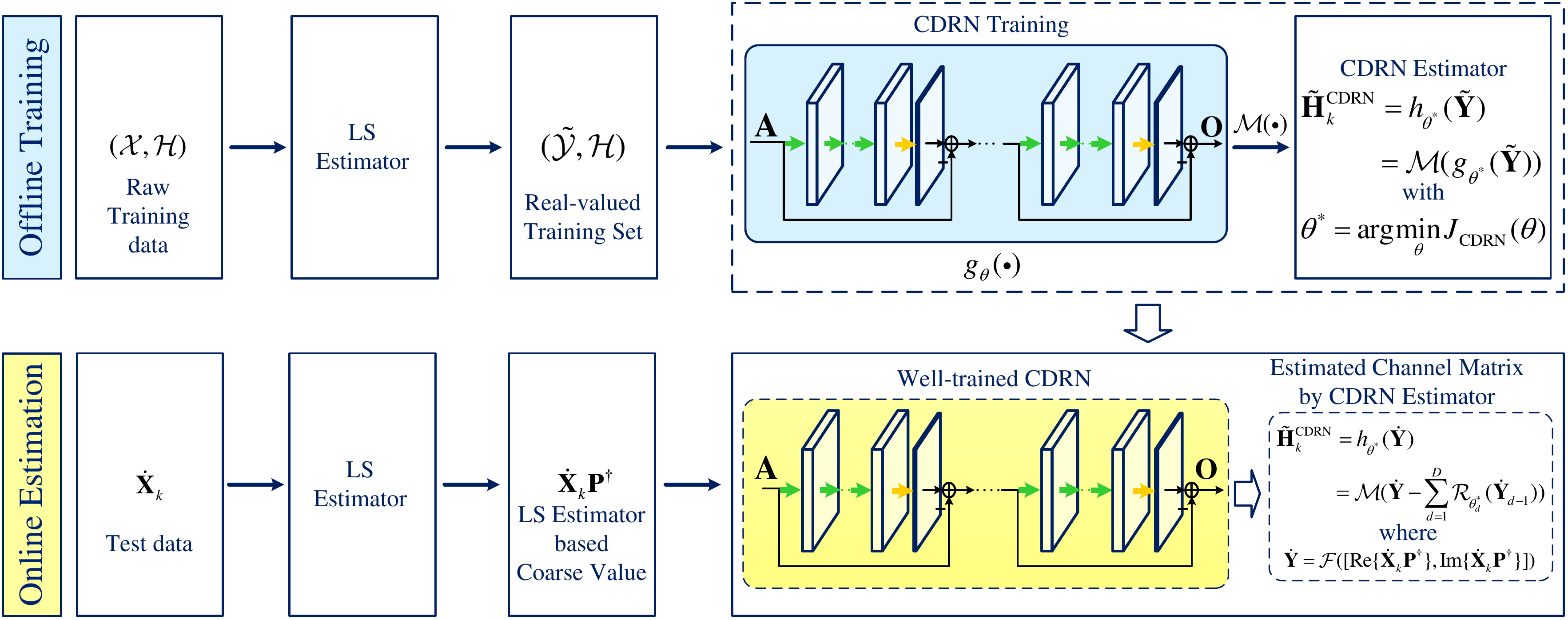} \vspace{-0.2 cm}
  \caption{ The developed CDRN-based framework for channel estimation in the considered IRS-MUC system. }\label{Fig:CDRN_framework}
  \vspace{-0.6 cm}
\end{figure}

\subsection{CDRN-based Channel Estimation Algorithm}
Based on the designed CDRN, we apply it to the developed DReL framework and propose a CDRN-based channel estimation algorithm, as shown in Fig. \ref{Fig:CDRN_framework}.
The proposed CDRL algorithm consists of an offline training and an online estimation.

\emph{a)~Offline Training of CDRN Algorithm}:

Based on the LS-based training set defined in (\ref{LS_trainingset}), we can obtain the following training set:
{\begin{equation}\label{LS_trainingset_CDRN}
  (\tilde{\mathcal{Y}},\mathcal{H}) = \{ ( \tilde{\mathbf{Y}}_k^{(1)},\mathbf{H}_k^{(1)} ), ( \tilde{\mathbf{Y}}_k^{(2)},\mathbf{H}_k^{(2)} ), \cdots, ( \tilde{\mathbf{Y}}_k^{(N_t)},\mathbf{H}_k^{(N_t)} )  \},
\end{equation}
where $\tilde{\mathbf{Y}}_k^{(i)} = \mathcal{F}([\mathrm{Re}\{\tilde{\mathbf{X}}_k^{(i)}\},\mathrm{Im}\{\tilde{\mathbf{X}}_k^{(i)}\}]) \in \mathbb{R}^{M \times (N+1) \times 2}$ and $(\tilde{\mathbf{Y}}_k^{(i)},\mathbf{H}_k^{(i)})$ represents the $i$-th training example of $(\tilde{\mathcal{Y}},\mathcal{H})$.
According to (\ref{Jmse}) and (\ref{h_theta}), the cost function of CDRN can be expressed as
\begin{equation}\label{Jmse_CDRN}
  J_{\mathrm{CDRN}}(\theta) = \frac{1}{2N_t}\sum_{i=1}^{N_t} \left\|h_{\theta}(\tilde{\mathbf{Y}}_k^{(i)}) - \mathbf{H}_k^{(i)}\right\|_F^2.
\end{equation}
Here, $h_{\theta}(\tilde{\mathbf{Y}}^{(i)}) = \mathcal{M}(g_{\theta}(\tilde{\mathbf{Y}}^{(i)}))$ where $\mathcal{M}(\cdot)\!\!:\mathbb{R}^{M \times (N+1) \times 2} \mapsto \mathbb{C}^{M \times (N+1)}$ represents a function that builds the complex-valued matrix based on the real-valued matrices.
As shown in Fig. \ref{Fig:CDRN_framework}, we can then use the BP algorithm to progressively update the network parameters to finally obtain the well-trained CDRN, i.e., the CDRN estimator:
\begin{equation}\label{CDRN_estimator}
  \tilde{\mathbf{H}}_k^{\mathrm{CDRN}} = h_{\theta^*}(\tilde{\mathbf{Y}}) = \mathcal{M}(g_{\theta^*}(\tilde{\mathbf{Y}})),
\end{equation}
where $\tilde{\mathbf{Y}}$ denotes the input matrix and $g_{\theta^*}(\cdot)$ is the function expression of the well-trained CDRN with the well-trained network parameters $\theta^* = [\theta^*_1,\theta^*_2,\cdots,\theta^*_D]$ and $h_{\theta^*}(\cdot)$ is the complex-valued form of $g_{\theta^*}(\cdot)$.}

{\emph{b)~Online Estimation of CDRN Algorithm}:}

{Given a test data denoted by $\dot{\mathbf{X}}_k$, we first send it to the LS estimator to obtain a coarse channel estimation denoted by $\dot{\mathbf{X}}_k\mathbf{P}^\dag$, and then send it to the CDRN estimator to obtain a refined channel estimate matrix:
\begin{equation}\label{CDRN_test}
  \tilde{\mathbf{H}}_k^{\mathrm{CDRN}} = h_{\theta^*}(\dot{\mathbf{Y}}) = \mathcal{M}\left(\dot{\mathbf{Y}} - \sum\limits_{d = 1}^{D}\mathcal{R}_{\theta^*_d}\left(\dot{\mathbf{Y}}_{d-1}\right)\right),
\end{equation}
where $\dot{\mathbf{Y}} = \mathcal{F}([ \mathrm{Re}\{\dot{\mathbf{X}}_k\mathbf{P}^\dag\}, \mathrm{Im}\{\dot{\mathbf{X}}_k\mathbf{P}^\dag\} ])$, $\dot{\mathbf{Y}}_{0} = \dot{\mathbf{Y}}$, and $\dot{\mathbf{Y}}_{d-1}$ and $\dot{\mathbf{Y}}_{d}$ are the input and output of the $d$-th denoising block of CDRN as defined in (\ref{I_d}).}

\emph{c)~CDRN-based Channel Estimation Algorithm Steps}:

Based on the above analysis, we then summarize the proposed CDRN-based channel estimation algorithm in Algorithm 1, where we use $i_t$ and $I_t$ to denote the iteration index and the maximum iteration number, respectively.

\begin{table}[t]
\small
\centering
\begin{tabular}{l}
\toprule[1.8pt] \vspace{-0.8 cm}\\
\hspace{-0.1cm} \textbf{Algorithm 1} {CDRN-based Channel Estimation Algorithm}  \\
\toprule[1.8pt] \vspace{-0.8 cm}\\
\textbf{Initialization:} $i_t = 0$, real-valued training set $(\tilde{\mathcal{Y}},\mathcal{H})$ \\
\textbf{Offline Training:} \\
1:\hspace{0.75cm}\textbf{Input:} Training set $(\tilde{\mathcal{Y}},\mathcal{H})$\\
2:\hspace{1.1cm}\textbf{while} $i_t \leq I_t $ \textbf{do} \\
3:\hspace{1.6cm}Update $\theta$ by BP algorithm to minimize $J_{\mathrm{CDRN}}(\theta)$ \\
\hspace{1.8cm} $i_t = i_t + 1$  \\
4:\hspace{1.1cm}\textbf{end while} \\
5:\hspace{0.75cm}\textbf{Output}:  Well-trained ${h}_{\theta^*}( \cdot ) $\\
\textbf{Online Estimation:} \\
6:\hspace{0.6cm}\textbf{Input:} Test data $\dot{\mathbf{Y}} = \mathcal{F}([ \mathrm{Re}\{\dot{\mathbf{X}}_k\mathbf{P}^\dag\}, \mathrm{Im}\{\dot{\mathbf{X}}_k\mathbf{P}^\dag\} ])$ \\
7:\hspace{0.95cm}\textbf{do} Channel Estimation using CDRN estimator in (\ref{CDRN_estimator}) \\
8:\hspace{0.6cm}\textbf{Output:} $\tilde{\mathbf{H}}_k^{\mathrm{CDRN}} = h_{\theta^*}(\dot{\mathbf{Y}})$. \vspace{-0.1cm}\\
\bottomrule[1.8pt]
\end{tabular}
\vspace{-0.8 cm}
\end{table}

\subsection{Theoretical Analysis}
To offer more insight of the proposed estimation algorithm, we then analyze the proposed CDRN and characterize its properties theoretically.
Although it is difficult to analyze the neural network which consists of a large number of parameters, we formulate the proposed CDRN as an explicit mathematical function and analyze the output of the CDRN theoretically.

As shown in Fig. \ref{Fig:CDRN_architecture}, the CDRN involves $D$ identical denoising blocks and the $d$-th denoising block can be expressed as the subtraction between the input $\mathbf{A}_{d-1}$ and the output of the residual subnetwork $\mathcal{R}_{\theta_d}(\cdot)$.
For the convenience of analysis, a simplified real-valued channel matrix is considered as the network input and it can be easily extended to a complex-valued model via a similar approach as in \cite{kay1993fundamentals}. In this case, the input matrix becomes two-dimension matrix and the function $\mathcal{M}(\cdot)$ should be removed for (\ref{Jmse_CDRN})-(\ref{CDRN_test}).
{Note that $\mathcal{R}_{\theta_d}(\cdot)$ consists of $N_l$ convolutional layers and both BN and ReLU operations have been shown to be able to improve the network stability and the training speed \cite{goodfellow2016deep}, \cite{glorot2011deep}.
For simplicity, we mainly focus on the effect of the convolution and ReLU operations on the network output.
Since the convolution operation can be expressed as the production of two matrices \cite{goodfellow2016deep}, for each residual subnetwork, we have
\begin{equation}\label{R_d_analysis_v1}
\begin{aligned}
\mathcal{R}_{\theta_d}(\mathbf{A}_{d-1}) & = \mathbf{Q}_{N_l}f_R(\mathbf{Q}_{N_{l-1}} \cdots f_R(\mathbf{Q}_2f_R(\mathbf{Q}_1\mathbf{A}_{d-1}))\cdots) \\
& = \mathbf{A}_{d-1} \mathbf{A}_{d-1}^{\dag}\mathbf{Q}_{N_l}f_R(\mathbf{Q}_{N_{l-1}} \cdots f_R(\mathbf{Q}_2f_R(\mathbf{Q}_1\mathbf{A}_{d-1}))\cdots)  \\
& = \mathbf{A}_{d-1} \mathbf{W}_d.
\end{aligned}
\end{equation}
Here, $\mathbf{W}_d = \mathbf{A}_{d-1}^{\dag}\mathbf{Q}_{N_l}f_R(\mathbf{Q}_{N_{l-1}} \cdots f_R(\mathbf{Q}_2f_R(\mathbf{Q}_1\mathbf{A}_{d-1}))\cdots)$, where $\mathbf{A}_{d-1}^{\dag} = \mathbf{A}_{d-1}^H(\mathbf{A}_{d-1}\mathbf{A}_{d-1}^H)^{-1}$ is the pseudoinverse of $\mathbf{A}_{d-1}$ and $\mathbf{Q}_l$ represents the network weights at the $l$-th, $l \in \{1,2,\cdots,N_l\}$, layer.
In addition, $f_R(x) = \max(0,x)$ denotes the ReLU activation function.
Note that although the output of the $d$-th residual subnetwork can be expressed as the product of two matrices, the elements in $\mathbf{W}_d$ are obtained through the non-linear operation and the residual subnetwork is still a non-linear network.}
{Thus, the output of the $d$-th denoising block can be expressed as
\begin{align}\label{A_d_analysis}
  \mathbf{A}_d = \mathbf{A}_{d-1} - \mathcal{R}_{\theta_d}(\mathbf{A}_{d-1})
  = \mathbf{A}_{d-1}\tilde{\mathbf{W}}_d = \prod_{i=1}^{d}\mathbf{A}\tilde{\mathbf{W}}_{i},
\end{align}
where $\mathbf{A}_0 = \mathbf{A}$ as defined in (\ref{I_d}) and $\tilde{\mathbf{W}}_d = \mathbf{I}_{N+1} - \mathbf{W}_d, \forall d \in \{1,2,\cdots,D\}$.
Based on (\ref{A_d_analysis}), the equation (\ref{R_d_analysis_v1}) can be rewritten as
\begin{equation}\label{R_d_analysis_v2}
  \mathcal{R}_{\theta_d}(\mathbf{A}_{d-1}) = \prod_{i=1}^{d-1}\mathbf{A}\tilde{\mathbf{W}}_i\mathbf{W}_d.
\end{equation}
Thus, substituting (\ref{R_d_analysis_v2}) into (\ref{h_theta}), the expression of CDRN can be written as
\begin{equation}\label{}
  g_{\theta}(\mathbf{A}) = \mathbf{A} - \sum\limits_{d = 1}^{D}\mathcal{R}_{\theta_d}\left(\mathbf{A}_{d-1}\right)
  = \mathbf{A} - \mathbf{A}\mathbf{W}_1 -  \sum_{d=2}^{D}\prod_{i=1}^{d-1}\mathbf{A}\tilde{\mathbf{W}}_i\mathbf{W}_d
  = \mathbf{A} - \mathbf{A}\mathbf{W},
\end{equation}
where $\mathbf{W} = \mathbf{W}_1 + \mathbbm{1}_{\Omega}(D)\sum_{d=2}^{D}\prod_{i=1}^{d-1}\tilde{\mathbf{W}}_i\mathbf{W}_d$ with $\Omega = \{a|a\geq2, a\in \mathbb{Z}\}$.
Note that we adopt the coarse estimated value obtained by the LS method as the input of the proposed CDRN.
Therefore, through learning from a large amount of LS-based training examples, the proposed CDRN estimator can exploit distinguishable features for denoising the LS-based input, i.e., our proposed method can further improve the estimation accuracy compared with the LS method. Based on this, we have $\mathbf{A} = \tilde{\mathbf{H}}_k^{\mathrm{LS}}$. In this case, the CDRN estimator, i.e., the well-trained CDRN can be formulated as
\begin{equation}\label{H_CDRN_analysis}
  \tilde{\mathbf{H}}_k^{\mathrm{CDRN}} = g_{\theta^*}(\tilde{\mathbf{H}}_k^{\mathrm{LS}}) = \tilde{\mathbf{H}}_k^{\mathrm{LS}} - \tilde{\mathbf{H}}_k^{\mathrm{LS}}\mathbf{W}^* = \tilde{\mathbf{H}}_k^{\mathrm{LS}}\left(\mathbf{I}_{N+1} - \mathbf{W}^*\right),
\end{equation}
where $\mathbf{W}^* = \mathbf{W}^*_1 + \mathbbm{1}_{\Omega}(D)\sum_{d=1}^{D}\prod_{i=1}^{d-1}\tilde{\mathbf{W}}_i^*\mathbf{W}_d^*$ denotes the well-trained matrix obtained from a well-trained network with parameter $\theta^*$.
Thus, the proposed CDRN is a data-driven non-linear estimator which can exploit more distinguishable features from a large amount of training examples to further improve estimation accuracy. }
On the other hand, according to (\ref{LSv1}) and (\ref{LMMSE}), the LMMSE estimator can be rewritten as \cite{kay1993fundamentals}
\begin{equation}\label{H_LMMSE_analysis}
\begin{aligned}
\tilde{\mathbf{H}}_k^{\mathrm{LMMSE}} &= \mathbf{H}_k^{\mathrm{LS}}\left(\mathbf{R}_{\mathbf{H}_k} + \frac{M\sigma_z^2}{C}\mathbf{I}_{N+1}\right)^{-1}\mathbf{R}_{\mathbf{H}_k} \\
  &=\mathbf{H}_k^{\mathrm{LS}}\left(\mathbf{I}_{N+1} - \mathbf{R}_{\mathbf{H}_k}^{-1}\left( \mathbf{R}_{\mathbf{H}_k}^{-1} + \frac{C}{M\sigma_z^2}\mathbf{I}_{N+1}\right)^{-1}\right).
\end{aligned}
\end{equation}
Based on (\ref{H_CDRN_analysis}) and (\ref{H_LMMSE_analysis}), the proposed CDRN estimator is equivalent to the LMMSE estimator when $\mathbf{W}^* = \mathbf{R}_{\mathbf{H}_k}^{-1}\left( \mathbf{R}_{\mathbf{H}_k}^{-1} + \frac{C}{M\sigma_z^2}\mathbf{I}_{N+1}\right)^{-1}$ is adopted.
That is, the proposed CDRN estimator is a generalized framework which subsumes an LMMSE estimator in a data driven manner as a subcase.

{According to the universal approximation theorem \cite{hornik1989multilayer}, the neural network has powerful capability in algorithmic learning and has been proven to be a universal function approximator.
Therefore, based on (\ref{H_CDRN_analysis}), (\ref{H_LMMSE_analysis}), and the universal approximation theorem, the proposed CDRN estimator is capable of approximating the LMMSE estimator through exploiting distinguishable features from a large amount of training examples, i.e., the proposed CDRN estimator can achieve the same performance as that of the LMMSE estimator.
Specifically, since the optimal MMSE estimator behaves the same as the LMMSE estimator when the prior PDF of $\mathbf{H}_k$ is a Gaussian PDF, the proposed CDRN estimator may achieve the optimal performance for the case of a Rayleigh fading channel.}
On the other hand, when the prior PDF of $\mathbf{H}_k$ is not a Gaussian PDF, the optimal MMSE estimator does not admit a closed-form. Then, the proposed CDRN estimator can exploit more distinguishable features from a large amount of training examples to achieve satisfactory estimation performance, while the LMMSE estimator can only obtain the limited estimation performance which is merely based on a second order channel correlation matrix.
These observations will be verified through simulations in the following section.

\begin{table}[t]
\normalsize
\caption{Simulation Settings}
\vspace{-0.3cm}
\centering
\small
\renewcommand{\arraystretch}{0.95}
\begin{tabular}{c c}
  \hline
  \vspace{-0.6cm} \\
   \textbf{Parameters} & \hspace{0.4cm} \textbf{Default Values} \hspace{0.4cm} \\
  \hline
  \vspace{-0.6cm} \\
  \vspace{-0.6cm} \\
   Path loss exponent of $U_k$-BS & $\gamma_1 = 3.6$  \\
  \vspace{-0.6cm} \\
  \vspace{-0.6cm} \\
   Path loss exponent of IRS-BS & $\gamma_2 = 2.3$ \\
  \vspace{-0.6cm} \\
  \vspace{-0.6cm} \\
   Path loss exponent of $U_k$-IRS & $\gamma_3 = 2$ \\
  \vspace{-0.6cm} \\
  \vspace{-0.6cm} \\
   Rician factor of $U_k$-BS & $\beta_{\mathrm{UB}} = 0$ \\
  \vspace{-0.6cm} \\
  \vspace{-0.6cm} \\
   Rician factor of IRS-BS & $\beta_{\mathrm{IB}} = 10$ \\
  \vspace{-0.6cm} \\
  \vspace{-0.6cm} \\
   Rician factor of $U_k$-IRS & $\beta_{\mathrm{UI}} = 0$ \\
  \vspace{-0.6cm} \\
  \hline
\end{tabular}
\vspace{-1cm}
\end{table}

\section{Numerical Results}
In this section, extensive simulations are provided to verify the efficientness of the proposed algorithm.
In the simulations, an IRS-MUC system consists of a BS equipped with $M$ antennas, an IRS with $N$ reflecting elements, and $K$ single-antenna users is considered, as defined in Fig. \ref{Fig:uplink scenario}.
Unless further specified, the IRS operates with continuous phase shifts and the default settings of the considered IRS-MUC system are $M=8$, $N=32$, $K=6$, and $C=N+1=33$.
A path loss model is adopted in the simulations and the path losses of each channel can be modeled as $\alpha_k^{\mathrm{UB}} = \alpha_0(\lambda_k^{\mathrm{UB}}/\lambda_0)^{-\gamma_1}$, $\alpha_k^{\mathrm{UI}} = \alpha_0(\lambda_k^{\mathrm{UI}}/\lambda_0)^{-\gamma_2}$, and $\alpha^{\mathrm{IB}} = \alpha_0(\lambda^{\mathrm{IB}}/\lambda_0)^{-\gamma_3}$, where $\gamma_i$, $i \in \{1,2,3\}$, is the path loss exponent, $\lambda_0 = 10~\mathrm{m}$ is the reference distance, $\alpha_0 = -15~\mathrm{dB}$ is the path loss at the reference distance, and $\lambda_k^{\mathrm{UB}}$, $\lambda_k^{\mathrm{UI}}$, and $\lambda^{\mathrm{IB}}$ represent the distances of $U_k$-BS, $U_k$-IRS, and IRS-BS, respectively.
Specifically, we set $\lambda_k^{\mathrm{UB}} = 100$ m, $\lambda_k^{\mathrm{UI}} = 16$ m, and $\lambda^{\mathrm{IB}} = 90$ m.
In addition, a Rician fading model is considered for all the channels to characterize the small-scale fading.
In this case, the channel of IRS-BS link can be expressed as
\begin{equation}\label{G_model}
  \mathbf{G} = \sqrt{\frac{\beta_{\mathrm{IB}}}{\beta_{\mathrm{IB}} + 1}}\mathbf{G}_{\mathrm{LOS}} + \sqrt{\frac{1}{\beta_{\mathrm{IB}} + 1}}\mathbf{G}_{\mathrm{NLOS}},
\end{equation}
where $\beta_{\mathrm{IB}}$ represents the Rician factor of IRS-BS channel, and $\mathbf{G}_{\mathrm{LOS}}$ and $\mathbf{G}_{\mathrm{NLOS}}$ are the line-of-sight (LOS) component and the Rayleigh component, respectively.
Note that the above model expression is a general model which involves both the LOS channel model (when $\beta_{\mathrm{IB}}\to\infty$) and the Rayleigh channel model (when $\beta_{\mathrm{IB}}=0$).
The channels of BS-$U_k$ and IRS-$U_k$ are generated similarly as defined in (\ref{G_model}). Similarly, we define the Rician factors of channels of $U_k$-BS and $U_k$-IRS as $\beta_{\mathrm{UB}}$ and $\beta_{\mathrm{UI}}$, respectively.
To consider the impacts of path loss factors, the elements in $\mathbf{G}$ should then be multiplied by $\sqrt{\alpha^{\mathrm{IB}}}$.
The detailed settings of the simulations are summarized in Table \Rmnum{2}, where the settings of $U_k$-BS link, i.e., $\beta_{\mathrm{UB}}=0$ and $\gamma_1 = 3.6$, are due to the rich scattering environment and the relatively large distance between $U_k$ and the BS.
Specifically, a normalized MSE (NMSE) is adopted as the estimation performance metric:
$  \mathrm{NMSE} = \frac{E( \|\tilde{\mathbf{H}} - \mathbf{H}\|_F^2 )}{E( \|\mathbf{H}\|_F^2 )}$
where $\tilde{\mathbf{H}}$ and $\mathbf{H}$ denote the estimated channel and the ground truth of the channel, respectively.
In addition, the signal-to-noise ratio (SNR) used in the simulation results refers to the transmit SNR, i.e., $\mathrm{SNR} = \mathcal{P}/\sigma_z^2$. {To evaluate the estimation performance, we also provide the simulation of the following algorithms for comparisons: the optimal MMSE method \cite{kay1993fundamentals}, the LMMSE method \cite{kay1993fundamentals}, the LS method \cite{biguesh2006training}, the bilinear alternating least squares (BALS) method \cite{de2020parafac}, and the binary reflection controlled LMMSE (B-LMMSE) method.} For the B-LMMSE method, we exploit the binary reflection controlled strategy proposed in \cite{mishra2019channel} and then adopt the LMMSE estimator for channel estimation.
In addition, a CDRN equipped with $D=3$ denoising blocks is adopted for our proposed method and the hyperparameters of CDRN are set according to Table \Rmnum{1}.
Each simulation point of the presented results is obtained by averaging the performance over $100,000$ Monte Carlo realizations{\footnotemark}\footnotetext{{The source code is available at https://github.com/XML124/CDRN-channel-estimation-IRS.}}.

Next, we will verify the channel estimation performance of the proposed CDRN method in terms of NMSE and unveil its relationship with SNR and channel dimensionality, respectively. Besides, visualizations of our proposed CDRN are provided to illustrate the denoising process.
Finally, the complexity analysis is also investigated to evaluate the computational complexity.

\begin{figure}[t]
  \centering
  \includegraphics[width=3.1in,height=2.8in]{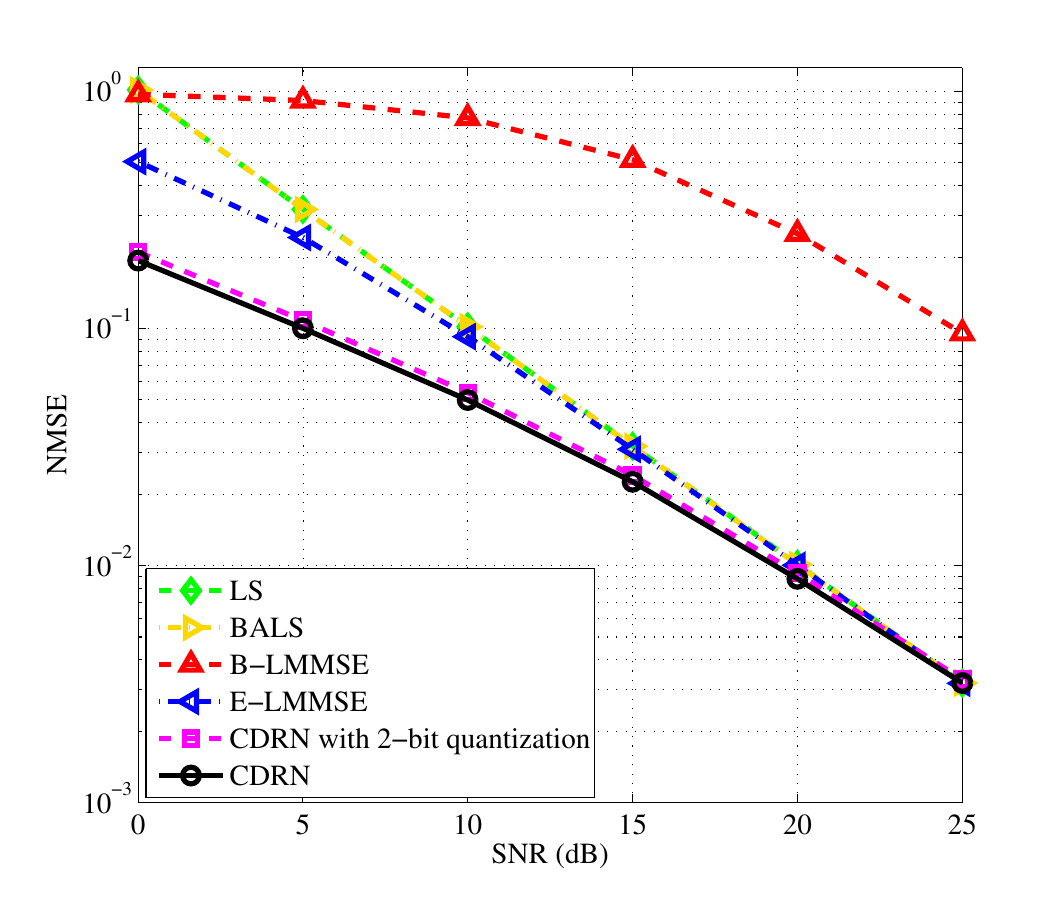}\vspace{-0.8cm}
  \caption{NMSE performance with different SNRs under $M$ = 8, $N = 32$, and $C = 33$.}\label{Fig:NMSE_SNR}\vspace{-1cm}
\end{figure}

\subsection{NMSE versus SNR}
The results of NMSE under different SNRs are presented in Fig. \ref{Fig:NMSE_SNR}. Since the cascaded channel does not follow the Rayleigh fading model, it is generally intractable to design the optimal MMSE estimator and derive the closed-form of LMMSE estimator. Hence, only an empirical LMMSE (denoted by E-LMMSE) can be obtained where the required statistical channel correlation matrix is replaced by a Monte Carlo based empirical channel correlation matrix.
As expected, the NMSEs of all the algorithms decrease with the increase of SNR since a higher transmit power can effectively mitigate the impact of noise.
Now, we first investigate the estimation performance in terms of different IRS control strategies. It can be seen that the DFT controlled methods (denoted by LS, E-LMMSE, and CDRN) outperform the B-LMMSE method significantly, especially under the high SNR regions. This is because for each time slot, all the $N$ elements are switched on for reflecting the incident signals when adopting the DFT controlled strategy, while only one element is switched on under the binary controlled strategy, leading to insufficient received signal power for channel estimation.
We then discuss the performance of the schemes adopting the DFT controlled methods. It is noticed that the LS method has a performance gap compared with the E-LMMSE method since the E-LMMSE can capture the statistical knowledge of the channel, while the LS method is designed by treating the channel as a deterministic but unknown constant.
Compared with LMMSE, the proposed CDRN method further improves the estimation performance and thus achieves the best performance among all the considered algorithms.
For example, the proposed CDRN method achieves a SNR gain of $5$ dB in terms of the NMSE $\approx 10^{-1}$ compared with the E-LMMSE method. The reason is that the E-LMMSE method is developed based on the constrained linear estimator.
In contrast, the proposed method adopts the non-linear CDRN to recover the channels by intelligently exploiting the spatial features of channels in a data driven approach.
{Besides, it can be observed from Fig. \ref{Fig:NMSE_SNR} that our proposed method can achieve a better NMSE performance compared with the BALS method. This is because the BALS method does not explore the statistical features of channels. In contrast, our proposed method can further improve the estimation performance by exploiting the channel statistical features from a large amount of LS-based training examples.}
On the other hand, low-quality phase shifters with a small number of quantization levels may be deployed to reduce the implementation cost \cite{wu2020towards, wu2019intelligent, you2020intelligent}. Therefore, it is necessary to investigate if the proposed channel estimation scheme is robust against the quantization of IRS phase shifts.
As shown in Fig. 7, the proposed CDRN method with a 2-bit uniformly quantized IRS phase shift \cite{wu2020towards, you2020intelligent} (denoted by CDRN with 2-bit quantization) almost behaves the same as the CDRN method with a continuous IRS phase shift.
This is because through offline training, the proposed CDRN can learn distinguishable features to address the introduced noise terms by the phase quantization errors.

\begin{figure}[t]
  \centering
  \includegraphics[width=4.1in,height=2.8in]{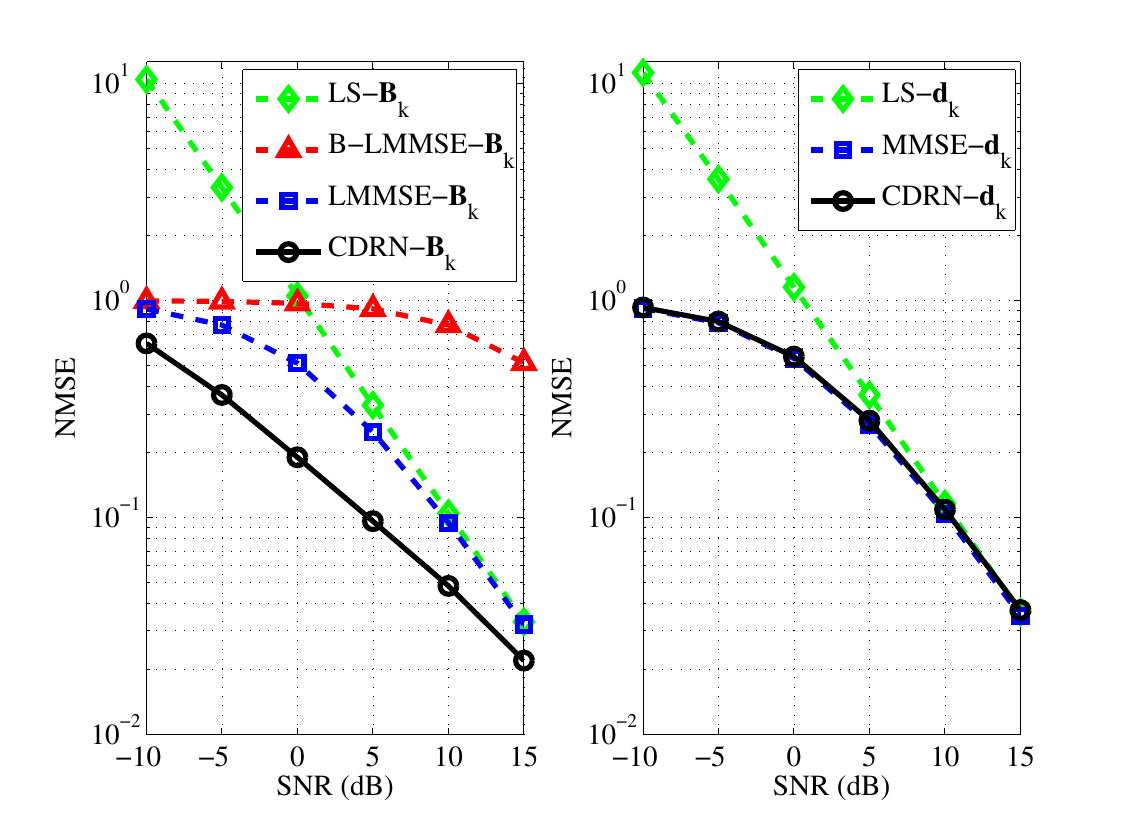} \vspace{-0.6cm}\\
  \hspace{0.2cm}{\scriptsize{(a) $\mathbf{B}_k$ for reflecting link. }} \hspace{2.3cm}{\scriptsize{(b) $\mathbf{d}_k$ for direct link. }} \hspace{0.3cm}\vspace{-0.4cm}\\
  \caption{NMSE performance with different SNRs under $M$ = 8, $N = 32$, and $C = 33$.}\label{Fig:NMSE_MC_SNR}\vspace{-1cm}
\end{figure}

To further evaluate the estimation performance, we present the NMSE results of the direct channel $\mathbf{B}_k$ and the reflection channel $\mathbf{d}_k$ in Fig. \ref{Fig:NMSE_MC_SNR}(a) and Fig. \ref{Fig:NMSE_MC_SNR}(b), respectively.
Similar to the results in Fig. \ref{Fig:NMSE_SNR}, Fig. \ref{Fig:NMSE_MC_SNR}(a) shows that the proposed CDRN still achieves the best performance in terms of the channel estimation of the reflection channel $\mathbf{B}_k$.
Since the direct channel follows Rayleigh fading model, the optimal MMSE method and the LMMSE method achieve the same performance as they have the same mathematical expression as in (\ref{LMMSE}).
Therefore, we compare three methods in Fig. \ref{Fig:NMSE_MC_SNR}(b): the LS method, the optimal MMSE method, and the proposed CDRN method.
It is shown that the CDRN method outperforms the LS method and achieves almost the same estimation performance as that of the optimal MMSE method, which presents the optimality of the proposed method.

\subsection{NMSE versus Channel Dimensionality}
\begin{figure}[t]
  \centering
  \includegraphics[width=3.1in,height=2.8in]{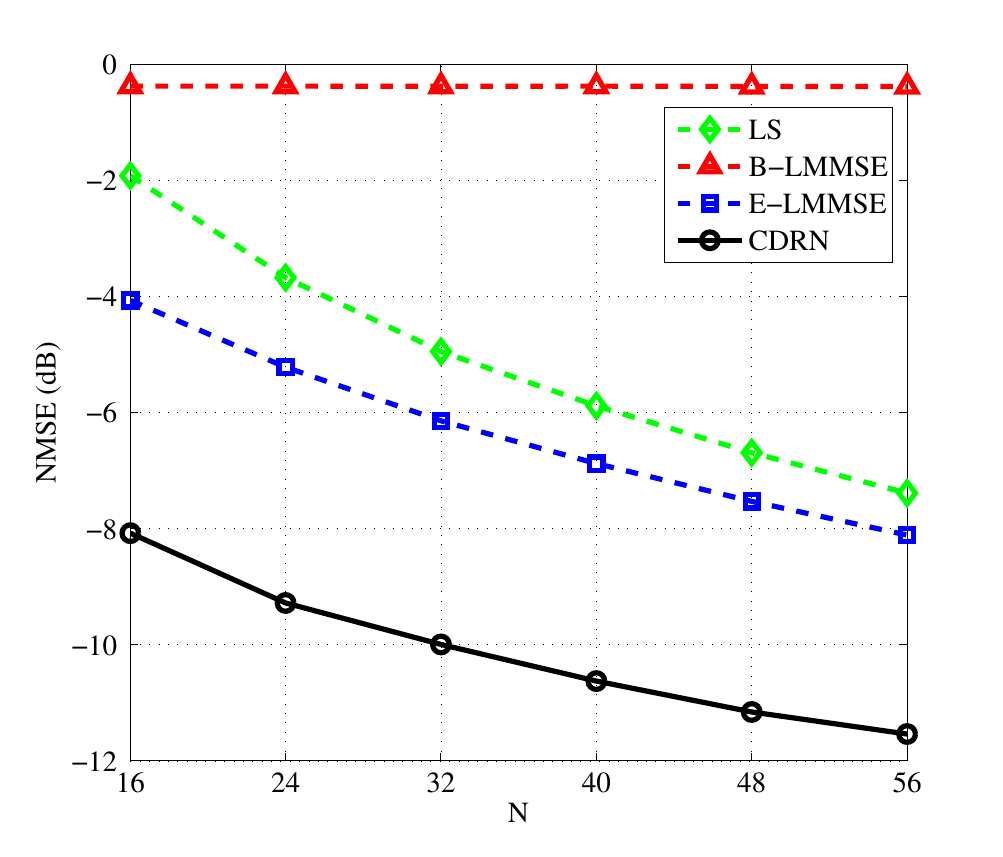}\vspace{-0.8cm}
  \caption{NMSE performance with different numbers of reflecting elements of IRS under $M$ = 8, $C = N+1$, and SNR = 5 dB.}\label{Fig:NMSE_N}\vspace{-0.8cm}
\end{figure}

\begin{figure}[t]
  \centering
  \includegraphics[width=3.1in,height=2.8in]{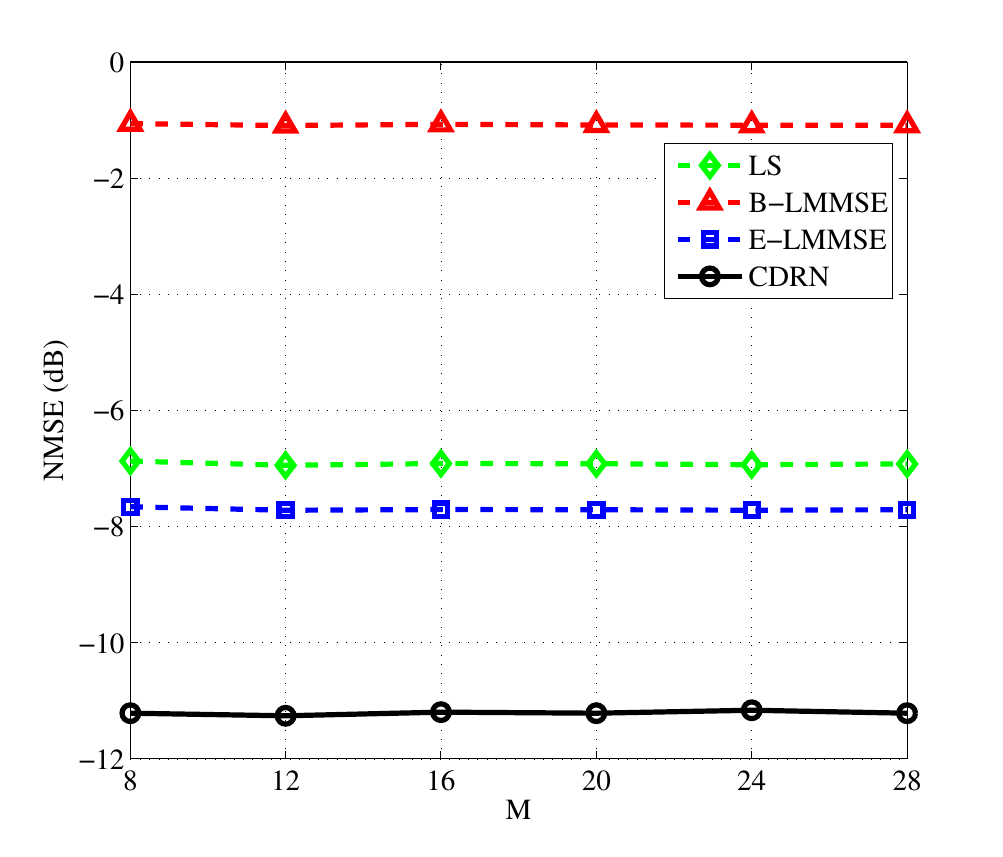}\vspace{-0.8cm}
  \caption{NMSE performance with different numbers of antennas at the BS under $N = 16$, $C = 17$, and SNR = 10 dB.}\label{Fig:NMSE_M}\vspace{-0.6cm}
\end{figure}

Note that the channel dimensionality is also an important parameter which affects the estimation performance. To illustrate the scalability of the proposed CDRN, we conduct simulations to evaluate the estimation performance with different channel dimensionality.
Specifically, the number of elements of IRS, denoted by $N$, directly affects the number of columns of channel matrix $\mathbf{H}_k$ and thus Fig. \ref{Fig:NMSE_N} plots the curves of NMSE versus $N$ for different algorithms.
It can be seen that the NMSE of DFT controlled methods (denoted by LS, E-LMMSE, and CDRN) decrease with the increase of $N$, while the NMSE binary controlled strategy based method, i.e., the B-MMSE method, remains almost a constant value for different $N$. This is because the B-MMSE method only switches on one element each time slot despite the total number of reflecting elements at the IRS.
In addition, the proposed CDRN method still achieves the best performance among all the considered algorithms since it is able to exploit more distinguishable features of channels from a larger size of input channel matrix.
Besides $N$, the number of antennas at the BS, denoted by $M$, determines the number of rows of $\mathbf{H}_k$ which also plays an important role in estimation performance. Therefore, Fig. \ref{Fig:NMSE_M} investigates the effect of $M$ on the NMSE performance.
It is obvious that the NMSE is insensitive to the change of $M$. This is because in the metric of NMSE, the array gain due to the increase of $M$ achieved in the numerator is neutralized by the one in the denominator.


\begin{figure}[t]
  \centering
  \includegraphics[width=3.1in,height=2.8in]{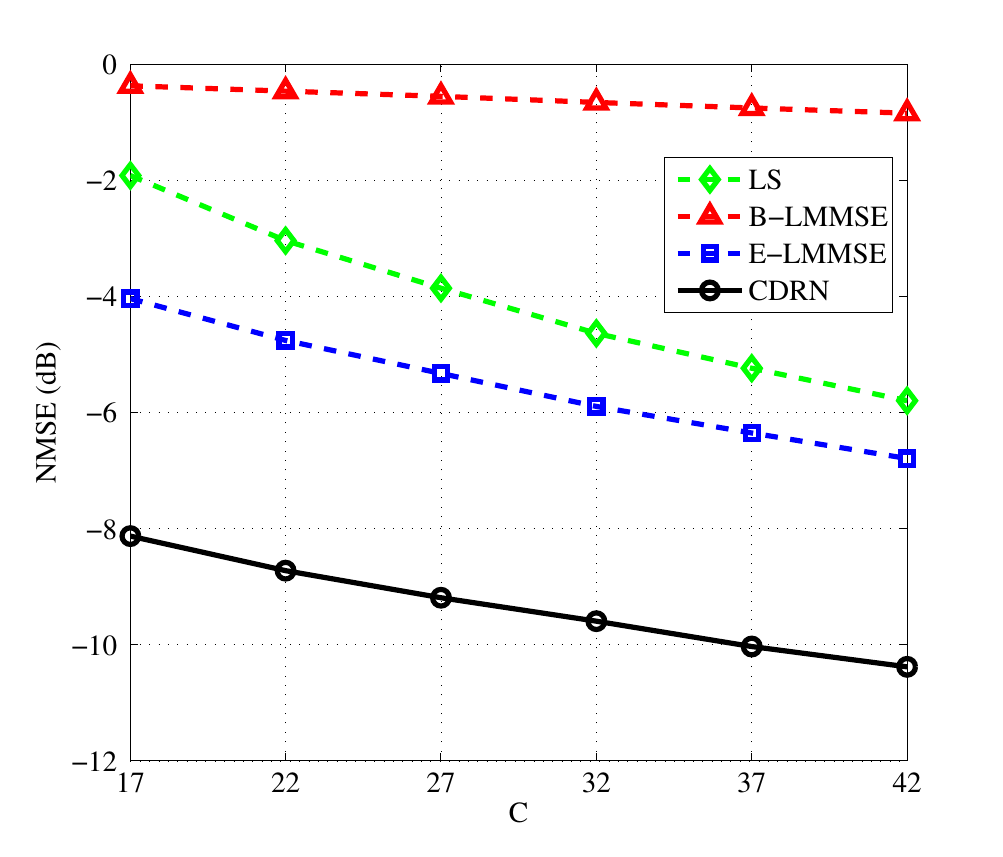}\vspace{-0.8cm}
  \caption{NMSE performance with different numbers of pilots under $M = 8$, $N = 16$, and SNR = 5 dB.}\label{Fig:NMSE_C}\vspace{-0.6cm}
\end{figure}

Moreover, the number of pilots $C\geq N+1$ has also impacts on the NMSE.
It is shown in Fig. \ref{Fig:NMSE_C} that the NMSEs of all the algorithms decrease with the increase of the number of pilots and different methods have different slopes with respect to $C$.
In particular, the NMSE of B-LMMSE method decreases gently with the increase of $C$, while the NMSEs of LS, E-LMMSE, and CDRN methods decrease sharply.
The reason is that for the B-LMMSE method, the power of the received signal at the BS is very limited which is only $1/N$ times of that of the other methods.
In addition, the proposed CDRN could exploit more distinguishable spatial features from a
larger scale input matrix to enhance the estimation accuracy, and thus achieves the best performance among all the considered algorithms, as presented in Fig. \ref{Fig:NMSE_C}.

\subsection{Visualization of CDRN}
To offer more insights of our proposed CDRN, we present the visualization of the outputs of each denoising block of a well-trained CDRN.
For the convenience of understanding, we only present the denoising process of ten points of a channel matrix, i.e., only these ten points are the received signals at the BS while the remaining points are the noise samples.
For visualization, we transform the input matrix and the output matrices of each denoising block to the grayscale images.
In particular, the elements of each matrix are normalized to an interval of $[0,1]$, where $0$ and $1$ are represented by black color and white color, respectively.

\begin{figure}[t]
\centering
\subfigure[]{
\includegraphics[height=0.16\linewidth]{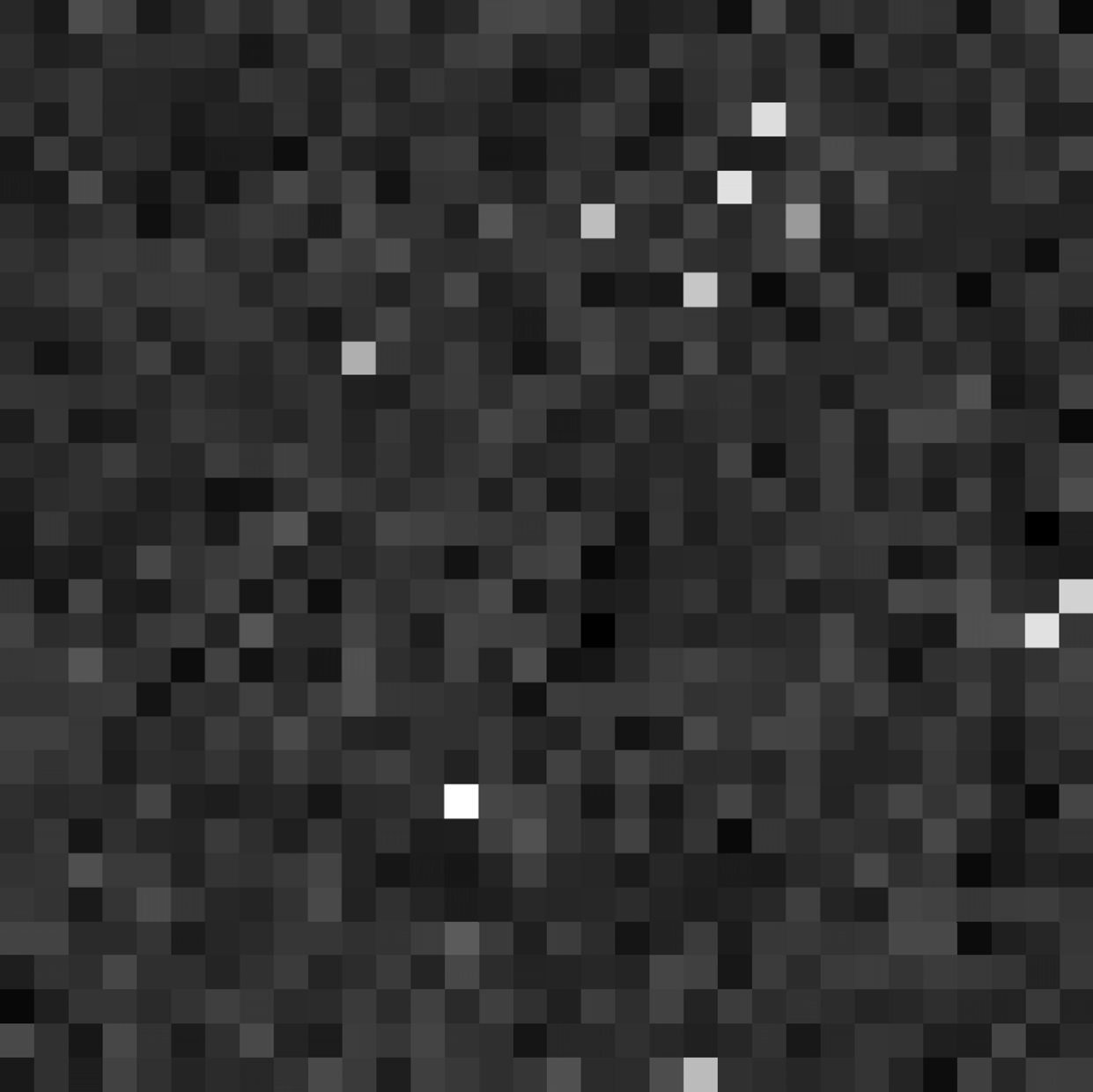}
\label{fig:Vis_Input}}
\,
\subfigure[]{
\includegraphics[height=0.16\linewidth]{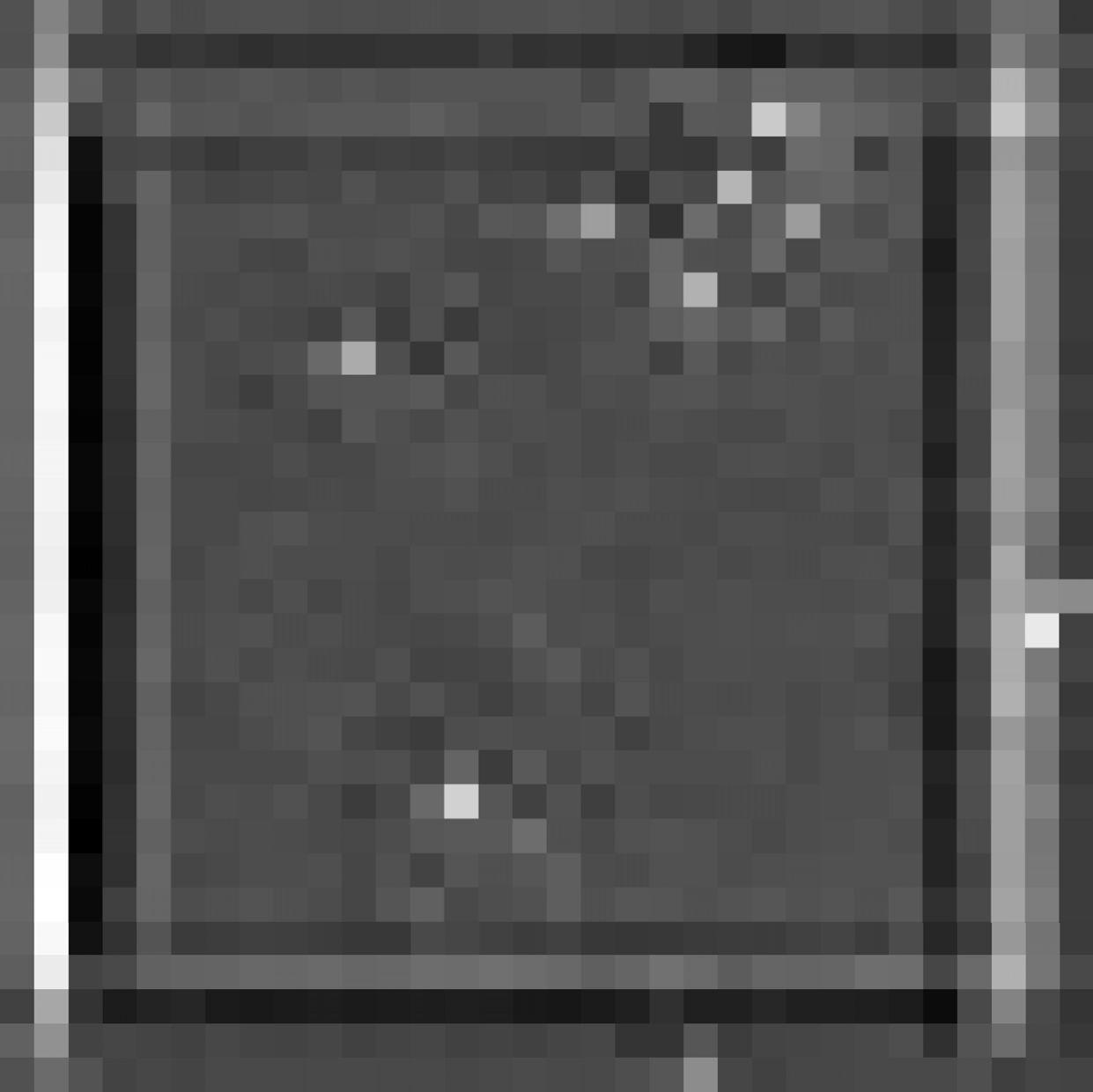}
\label{fig:Vis_B1}}
\,
\subfigure[]{
\includegraphics[height=0.16\linewidth]{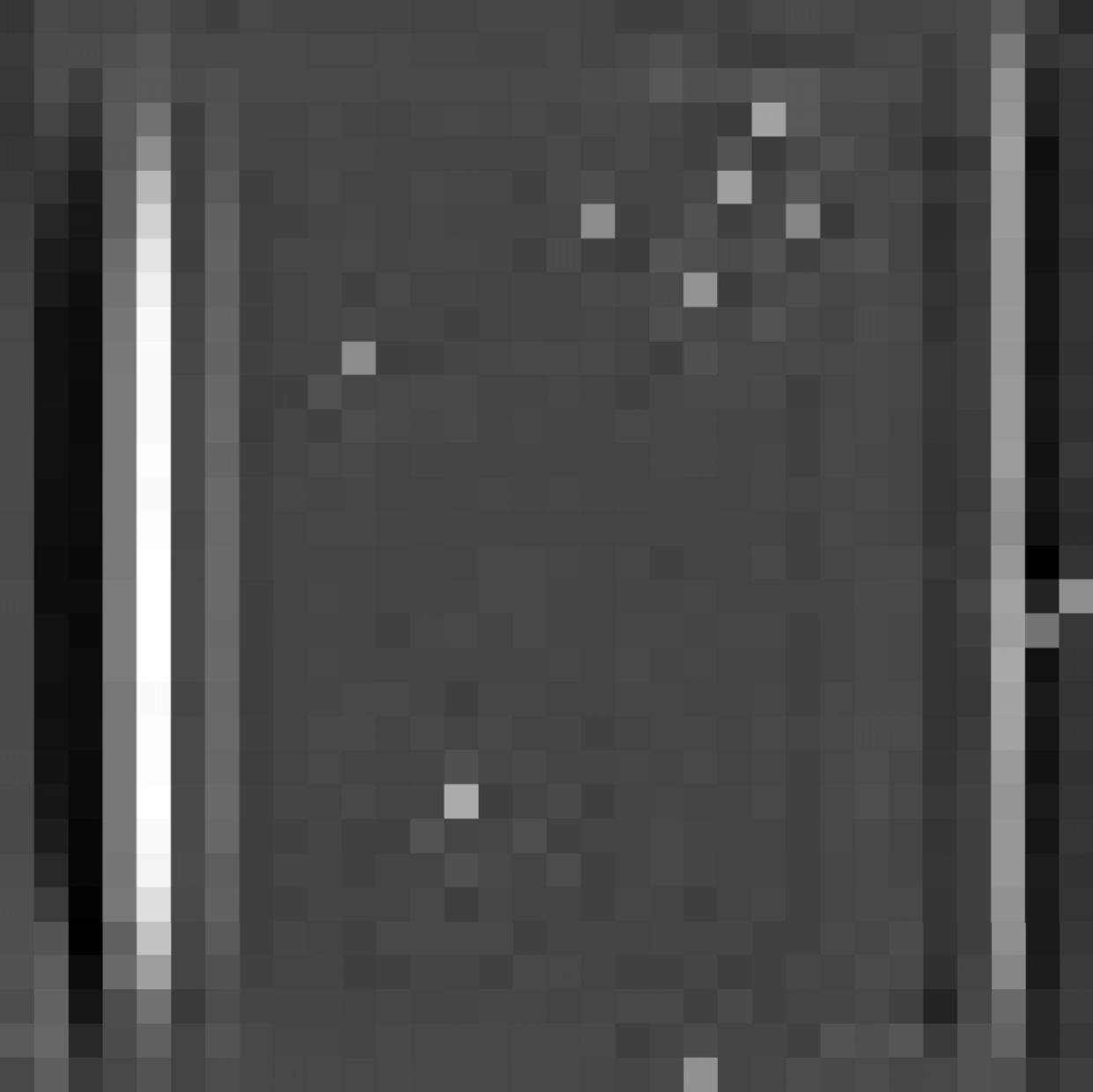}
\label{fig:Vis_B2}}
\,
\subfigure[]{
\includegraphics[height=0.16\linewidth]{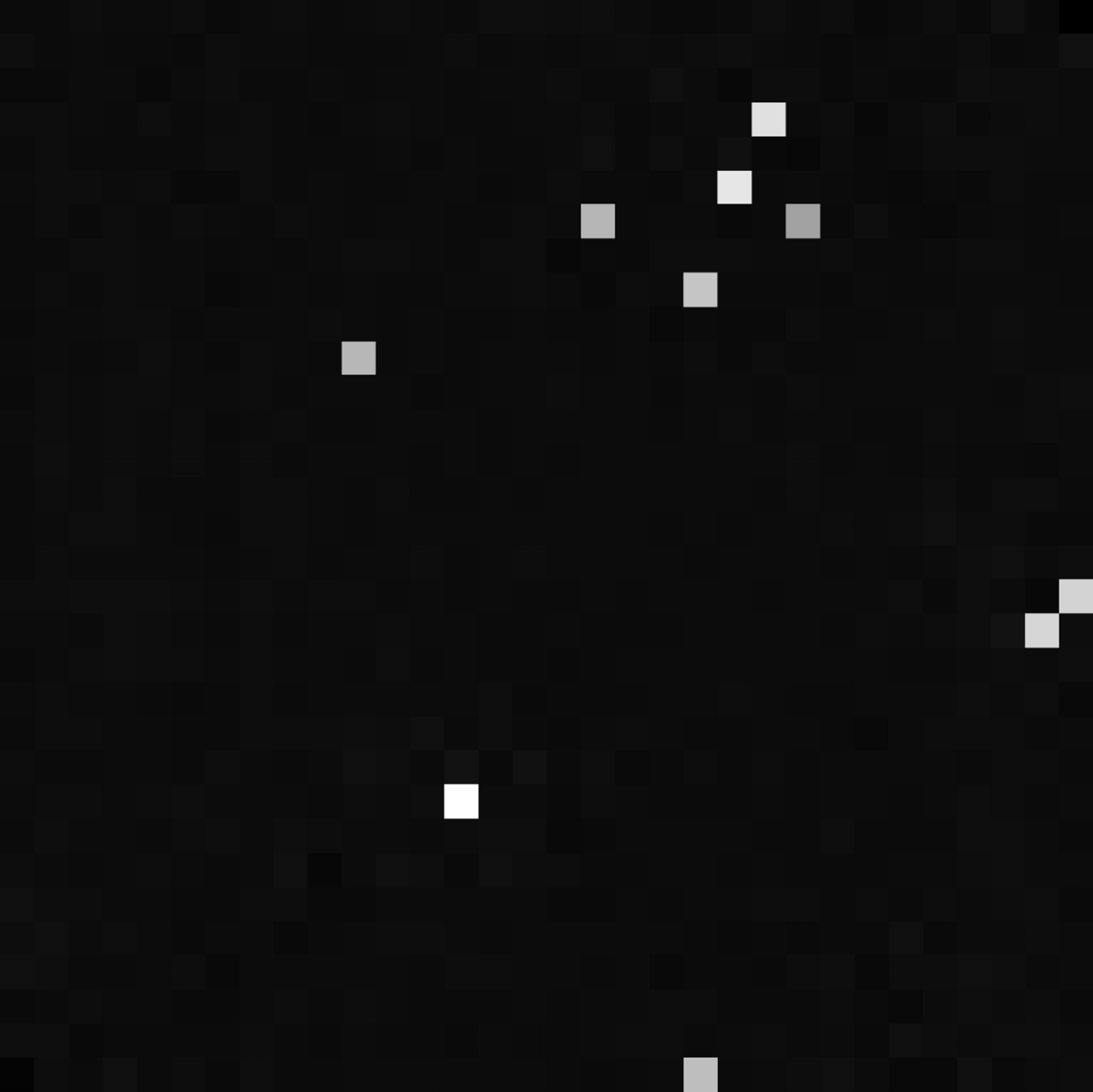}
\label{fig:Vis_B3}}
\vspace{-0.2 cm}
\caption{A visualization of denoising blocks of a well-trained CDRN with $D=3$ denoising blocks under SNR = 16 dB. (a) $\mathbf{A}$, the input of CDRN; (b) $\mathbf{A}_1$, the output of the first denoising block; (c) $\mathbf{A}_2$, the output of the second denoising block; (d) $\mathbf{A}_3$, the output of the last denoising block, i.e., the output of CDRN.}
\vspace{-0.8 cm}
\end{figure}

According to Fig. \ref{Fig:CDRN_architecture}, the input of CDRN, the output of the first, second, and last denoising blocks are represented by $\mathbf{A}$, $\mathbf{A}_1$, $\mathbf{A}_2$, and $\mathbf{A}_3$, respectively.
The grayscaled image of $\mathbf{A}$ is shown in Fig. \ref{fig:Vis_Input}, where the ten distinguishable bright pixels are the interest channel coefficients while the remaining pixels are the noise samples. The cluttered noise pixels indicate that the noise samples have a large variance.
Subsequently,  Fig. \ref{fig:Vis_B1} and Fig. \ref{fig:Vis_B2} are the visualizations of $\mathbf{A}_1$ and $\mathbf{A}_2$, where we find that the noise pixels become tidy, i.e., the noisy channel matrix are denoised progressively. Finally, the output of $\mathbf{A}_3$, i.e., the output of CDRN is visualized in Fig. \ref{fig:Vis_B3}, where the noise pixels are largely eliminated and thus a denoised channel matrix is obtained.
In fact, the proposed CDRN method adopts a stage-by-stage denoising mechanism and achieves a satisfactory performance. \vspace{-0.2 cm}

\subsection{Complexity Analysis}
Finally, we analyze the computational complexities of the considered algorithms and summarize the results in Table \ref{computational_complexity}.
Here, $n_0$ denotes the input dimension, $n_l$ and $s_l$ represent the number of neural network channels and the spatial size (length of side) of the filter of the $l$-th, $l \in \{1,2,\cdots, N_l\}$, convolutional layer, respectively.
In addition, $N_t$ and $I_t$ are both defined in Algorithm 1.
Specifically, both the B-LMMSE method and the E-LMMSE method have the same computational complexity, thus we use LMMSE to denote the B-LMMSE/E-LMMSE method.
Note that the computational complexities of the LS and LMMSE methods only arise from the online estimation, while the proposed CDRN method has an additional computational complexity due to the offline training.
{Specifically, the LS method can be realized via the inverse fast Fourier transform (IFFT).} While the computational cost of the LMMSE method mainly comes from the matrix inversions.
{As for the proposed method, its computational cost is dominated by the computations of the CDRN \cite{he2015convolutional} and the LS-based input. For simplicity, we only present the highest order terms of the complexities in Table III, as commonly adopted in e.g., \cite{arora2009computational}.} It can be seen from Table \ref{computational_complexity} that compared with the LS and LMMSE methods, the proposed CDRN method can achieve the best system performance at the expense of a higher computational complexity.
On the other hand, although the proposed CDRN involves a large number of parameters, the required actual online estimation time can be greatly reduced by exploiting the parallelization of a graphics processing unit (GPU) \cite{goodfellow2016deep}. This conclusion can be verified in Table \ref{computation_time}, where each result is obtained by executing the algorithms on a desktop computer with an i7-8700 3.20 GHz central processing unit (CPU) and a Nvidia GeForce RTX 2070 GPU.
Therefore, although the proposed CDRN method has a relatively higher computational complexity compared with the LS and LMMSE methods, the associated computation time can be greatly reduced through exploiting the parallel computing of GPU.
\vspace{-0.5cm}

\begin{table}[t]
\caption{Computational complexities of different estimation algorithms }
\vspace{-0.2 cm}
\label{computational_complexity}
\vspace{-0.2 cm}
\centering
\small
\renewcommand{\arraystretch}{1.25}
\begin{tabular}{c c c}
  \toprule\vspace{-1cm}\\
  {\textbf{\footnotesize{Algorithm}}} & {\textbf{\footnotesize{Online Estimation}}} & {\textbf{\footnotesize{Offline Training}}} \\
  \toprule
  {\textbf{\footnotesize{LS}}}
  & \footnotesize{{$O(MC\log_2(MC))$ }}
  & -
  \\
  \hline
  {\textbf{\footnotesize{LMMSE}}}
  & \footnotesize{$O(C^3+MC^2)$}
  & -
  \\
  \hline
  \vspace{-0.5cm} \\
  {{\textbf{\footnotesize{CDRN}}}}  & \footnotesize{$O\left( DM(N+1)\sum\limits_{l = 1}^{N_l}n_{l-1} \cdot s_l^2 \cdot n_l\right)$} &
  \footnotesize{$O\left( N_tI_tDM(N+1)\sum\limits_{l = 1}^{L}n_{l-1} \cdot s_l^2 \cdot n_l \right)$}
  \vspace{0.1cm}
  \\
  \toprule
\end{tabular}
\vspace{-0.4 cm}
\end{table}

\begin{table}[t]
\caption{Computation time (in seconds) of different estimation algorithms }
\vspace{-0.2 cm}
\label{computation_time}
\vspace{-0.2 cm}
\centering
\renewcommand{\arraystretch}{1.25}
\begin{tabular}{c c c}
  \toprule\vspace{-0.85cm}\\
  {\textbf{\footnotesize{Algorithm}}} & {\textbf{\footnotesize{Online Estimation}}} & {\textbf{\footnotesize{Offline Training}}} \\
  \toprule
  {\textbf{\footnotesize{LS}}}
  & \footnotesize{$1.26 \times 10^{-3}$}
  & -
  \\
  \hline
  {\textbf{\footnotesize{LMMSE}}}
  & \footnotesize{$4.56 \times 10^{-3}$}
  & -
  \\
  \hline
  {{\textbf{\footnotesize{CDRN}}}}
  & \footnotesize{$2.66 \times 10^{-3}$}
  & \footnotesize{$492.58$}
  \\
  \toprule
\end{tabular}
\vspace{-0.8 cm}
\end{table}

\section{Conclusion}
This paper proposed a data-driven DReL approach to address the channel estimation problem for IRS-MUC systems operating with TDD protocol.
Firstly, we formulated the channel estimation problem in IRS-MUC systems as a denoising problem and developed a versatile DReL-based channel estimation framework. Specifically, according to the Bayesian MMSE criterion, a DRN-based MMSE was derived and analyzed in terms of Bayesian philosophy.
Under the developed DReL-based framework, we then designed a CDRN to denoise the noisy channel matrix for channel recovery. For the proposed CDRN, a CNN-based denoising block with an element-wise subtraction structure was specifically designed to exploit both the spatial features of the noisy channel matrices and the additive nature of the noise simultaneously.
Taking advantages of the superiorities of CNN and DReL in feature extraction and denoising, the proposed CDRN estimation algorithm can further improve the estimation accuracy.
Simulation results demonstrated that the proposed method can achieve almost the same estimation accuracy as that of the optimal MMSE estimator based on a prior PDF of channel.

\bibliographystyle{ieeetr}

\setlength{\baselineskip}{10pt}

\bibliography{ReferenceSCI2}

\end{document}